\documentclass[aps,prl,twocolumn,showpacs,superscriptaddress,groupedaddress]{revtex4}  

\usepackage{graphicx}  
\usepackage{dcolumn}   
\usepackage{bm}        
\usepackage{amssymb}   

\begin{document}

\widetext

\title{Origin of fermion masses without spontaneous symmetry breaking}
\author{Venkitesh Ayyar}
\author{Shailesh Chandrasekharan}
\affiliation{Department of Physics, Box 90305, Duke University,
Durham, North Carolina 27708, USA}

\date{\today}

\begin{abstract}
Using a simple three dimensional lattice four-fermion model we argue that massless fermions can become massive due to interactions without the need for any spontaneous symmetry breaking. Using large scale Monte Carlo calculations within our model, we show that this non-traditional mass generation mechanism occurs at a second order quantum critical point that separates phases with the same symmetries. Universality then suggests that the new origin for the fermion mass should be of wide interest.
\end{abstract}

\pacs{71.10.Fd,02.70.Ss,11.30.Rd,05.30.Rt}

\maketitle

The origin of mass in the universe is an interesting problem in fundamental physics \cite{Wilczek:2012sb}. In continuum quantum field theory, fermion masses arise from local fermion bilinear mass terms that are introduced as parameters in the theory. If symmetries of the theory prevent such terms, perturbatively fermions remain massless. However, these symmetries can break spontaneously and form non-zero fermion bilinear condensates dynamically that can make fermions massive. This well-known mechanism of mass generation is used in the standard model of particle physics to give quarks and leptons their masses. Recent progress in the field of topological insulators suggests the existence of an alternate mechanism for the origin of the fermion mass \cite{Kitaev:2009mg,Fidkowski:2013jua,Gu:2012ib,Metlitski:2014xqa}. These studies show that fermions can become massive without spontaneous symmetry breaking due to an interplay of quantum entanglement and topology. In particular, fermion bilinear condensates are not necessary. An example of how the mechanism works in two space-time dimensions has been discussed recently \cite{BenTov:2014eea}. Since this alternate mechanism also constrains the number of fermion species of the theory, some speculations that it may explain the particle spectrum of the standard model have also been proposed \cite{Wen:2013ppa,You:2014oaa,You:2014vea}.

Quantum entaglement and topology can be important in determining the ground state properties of matter \cite{Chen:2011pg}.  They can also lead to exotic second order phase transitions that are characterized by a change in some topological order of the ground state, unlike standard transitions which occur due to spontaneous symmetry breaking characterized by a change in some local order parameter. In certain cases, fractionalization of the fundamental degrees of freedom can occur leading to emergent gauge fields \cite{Senthil05032004}. Understanding these exotic second order phase transitions is an active area of research today \cite{Isakov13012012,PhysRevB.86.155131}. It was recently proposed that the origin of fermion mass without spontaneous symmetry breaking can also be related to a similar exotic quantum phase transition. Evidence for this conjecture was provided using Monte Carlo calculations on small lattices in a model inspired by the physics of electrons hopping on a honeycomb lattice \cite{Slagle:2014vma,He:2015bda}. In this work we show a similar phenomenon in a simple lattice field theory model in three space-time dimensions in the action formalism that is much closer in spirit to particle physics. Due to the simplicity of the model we are able to perform calculations on a much larger scale and estimate the critical exponents. Some of the technical details of our work on small lattices have already appeared earlier in \cite{Ayyar:2014eua} and have also been verified recently in \cite{Catterall:2015zua}. 

It is possible to understand the alternative mechanism of fermion mass generation qualitatively using four-fermion models. Consider free massless fermions interacting through a four-fermion coupling that has symmetries which forbid the generation of fermion bilinear mass terms through radiative corrections. Since perturbatively four-fermion couplings are irrelevant, the model will contain a massless fermion phase at weak couplings. However at strong couplings such models can be in a massive phase. Traditionally this massive phase is accompanied with spontaneous symmetry breaking and the generation of fermion bilinear condensates. In fact gauge anomalies force the existence of such a phase due to t'Hoofts anomaly matching arguments \cite{tHooft:1979bh}. But, if the four-fermion couplings explicitly break all anomalous symmetries while still forbidding fermion bilinear condensates, then there may be a way to circumvent anomaly matching arguments and allow a massive fermion phase without any spontaneous symmetry breaking. 

\begin{figure}
\includegraphics[width=0.48\textwidth]{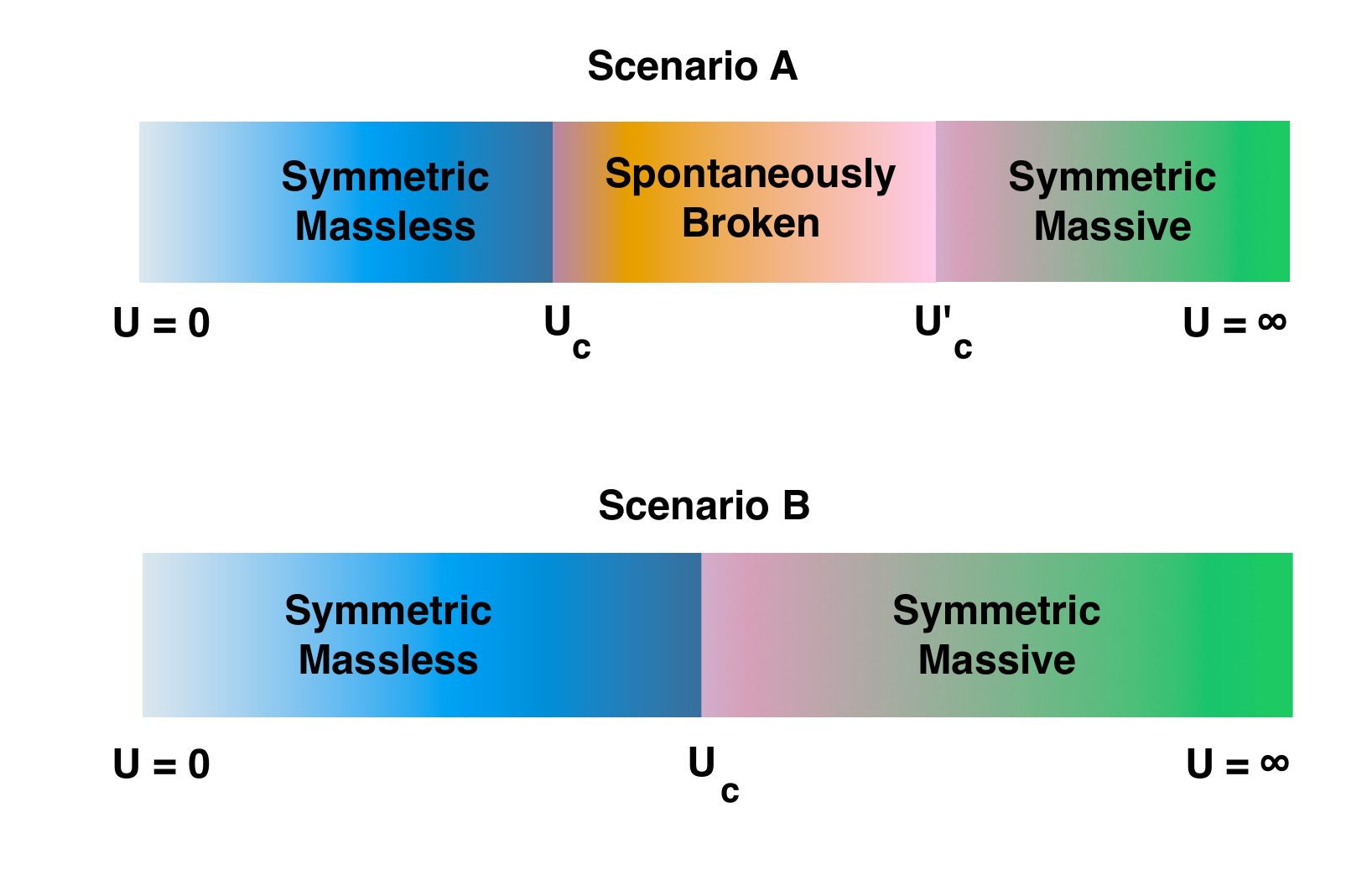}
\caption{\label{fig1} The two possible scenarios for the phase diagram of lattice four-fermion models that show the existence of a symmetric massless fermion phase at weak couplings and a symmetric massive fermion phase without spontaneous symmetry breaking at strong couplings. Previous studies in four space-time dimensions found results consistent with scenario A, where an intermediate spontaneously broken phase was found. Our work in three space-time dimensions is consistent with scenario B with a quantum critical point at $U_c$.}
\end{figure}

Lattice models that realize the above speculative scenario concretely were constructed and studied long ago \cite{Hasenfratz:1988vc,Hasenfratz:1989jr,Lee:1989mi,Lee:1989xq}. These models contained a massless fermion phase at weak couplings and a massive fermion phase at strong couplings. But most interestingly, the massive phase was not accompanied by spontaneous breaking of any symmetries. The two phases were usually separated by a more conventional spontaneously broken phase as shown in Fig.~\ref{fig1} as scenario A. The exotic massive fermion phase without spontaneous symmetry breaking can be interesting in continuum quantum field theory if it is connected to the massless fermion phase directly through a second order transition. This is shown as scenario B in Fig.~\ref{fig1}. The presence of a second order critical point will allow one to take the continuum limit of the massive fermion phase and thus eliminate the effects of lattice discretization. Most earlier work had focused on four Euclidean dimensions, and a direct second order transition was never discovered, although some mean field theory calculations predicted a direct first order transition \cite{Stephanov:1990sq,Stephanov:1990pc}. In our work we focus on three Euclidean dimensions and find strong evidence for a direct second order transition (scenario B in Fig.~\ref{fig1}).

The model we study contains four flavors of massless {\em reduced lattice staggered fermions} on a cubical space-time lattice with an onsite four-fermion interaction. Each lattice fermion flavor describes a single four-component Dirac fermion in the continuum due to fermion doubling \cite{Sharatchandra:1981si,vandenDoel:1983mf,Golterman:1984cy}.Our model can be obtained as a limit of lattice Yukawa models that were studied long ago in four space-time dimensions \cite{Lee:1989mi,Lee:1989xq}. Further, our model has the same fermion content as the honeycomb lattice models studied recently \cite{Slagle:2014vma,He:2015bda}. We use four-component Grassmann valued fields, $ \psi_{x,i},\ i = 1,2,3,4$, on each lattice site $x$ to describe the fermion fields. Then, the Euclidean action of our model is given by :
\begin{equation}
S = \sum_{i=1}^4\ \sum_{x,y} {\psi}_{x,i} \ M_{x,y} \ \psi_{y,i} - U \sum_x \psi_{x,1}\psi_{x,2}\psi_{x,3}\psi_{x,4}
\label{act}
\end{equation}
where $M$ is the well known massless staggered fermion matrix given by
\begin{equation}
M_{x,y} \ =\  \sum_{\hat{\alpha}=1,2,3} \frac{\eta_{x,{\hat{\alpha}}}}{4}\ [\delta_{x,y+\hat{\alpha}} - \delta_{x,y-\hat{\alpha}}],
\label{staggered}
\end{equation}
where $\eta_{x,\hat{\alpha}}$ are phases that introduce a $\pi$-flux through all plaquettes.
In our work we study cubical lattices of equal size $L$ in each direction with anti-periodic boundary conditions. Observables are defined as usual through the Grassmann integral
\begin{equation}
\langle {\cal O}\rangle = 
\frac{1}{Z} \int \ \Big(\prod_{i,x} [d\psi_{x,i}]\Big) \  {\cal O} \ \mathrm{e}^{-S}.
\end{equation}
where $Z$ is the partition function.

The action given in Eq.~(\ref{act}) is symmetric under the usual space-time lattice transformations and an internal $SU(4)$ flavor transformations \cite{Ayyar:2014eua}. Using weak coupling and strong coupling perturbation theory, it is easy to argue that all lattice symmetries remain unbroken at both weak and strong couplings. Thus, the essential question is whether there is a single transition between the two phases or is there an intermediate phase where some of the lattice symmetries are broken. Previous studies in four space-time dimensions do seem to find such an intermediate phase. Here we present clear evidence from large lattices for a single second order transition between the two phases in three space-time dimensions and estimate the critical exponents at the transition.

\begin{figure}
\includegraphics[width=0.48\textwidth]{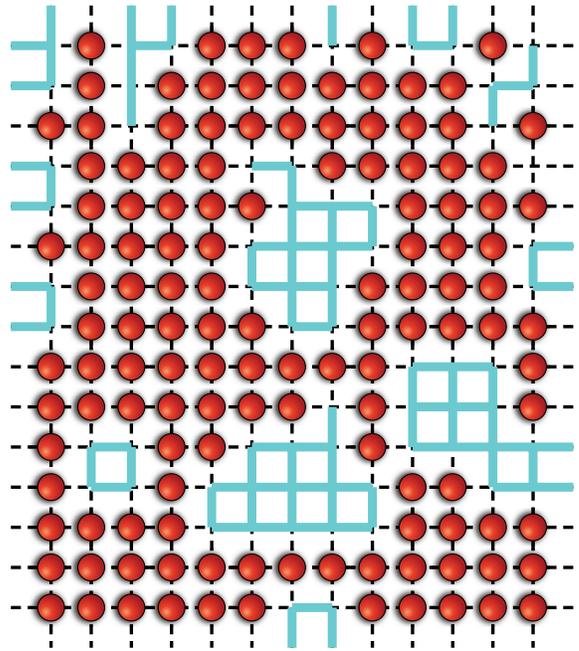}
\caption{\label{fig2} An example of a monomer configuration $[n]$ showing free fermion bags on a two dimensional lattice. The filled circles represent monomers and the connected regions without monomers form free fermion bags.}
\end{figure}

We perform calculations using the fermion bag approach \cite{PhysRevD.82.025007} where the problem is converted into a statistical mechanics of monomer configurations which we denote as $[n]$. Each monomer configuration is defined through a binary lattice field $n_x = 0,1$ which denotes the absence or presence of a monomer at the site $x$ respectively. Figure \ref{fig2} shows an illustration of a monomer configuration on a two dimensional lattice.  Each monomer represents a four-fermion interaction and free fermions hop on sites that do not contain monomers. The fermion bag approach also gives a very intuitive picture of the underlying physics: At small couplings the monomer density is small and fermions are essentially free, while at strong couplings the lattice is filled with monomers with very few empty sites for free fermions to hop making them massive. Since monomers form local singlets, both phases have the same lattice symmetries. Further details of our computational approach, including algorithms that we use can be found in \cite{Ayyar:2014eua}.

\begin{figure}
\vskip0.2in
\includegraphics[width=0.48\textwidth]{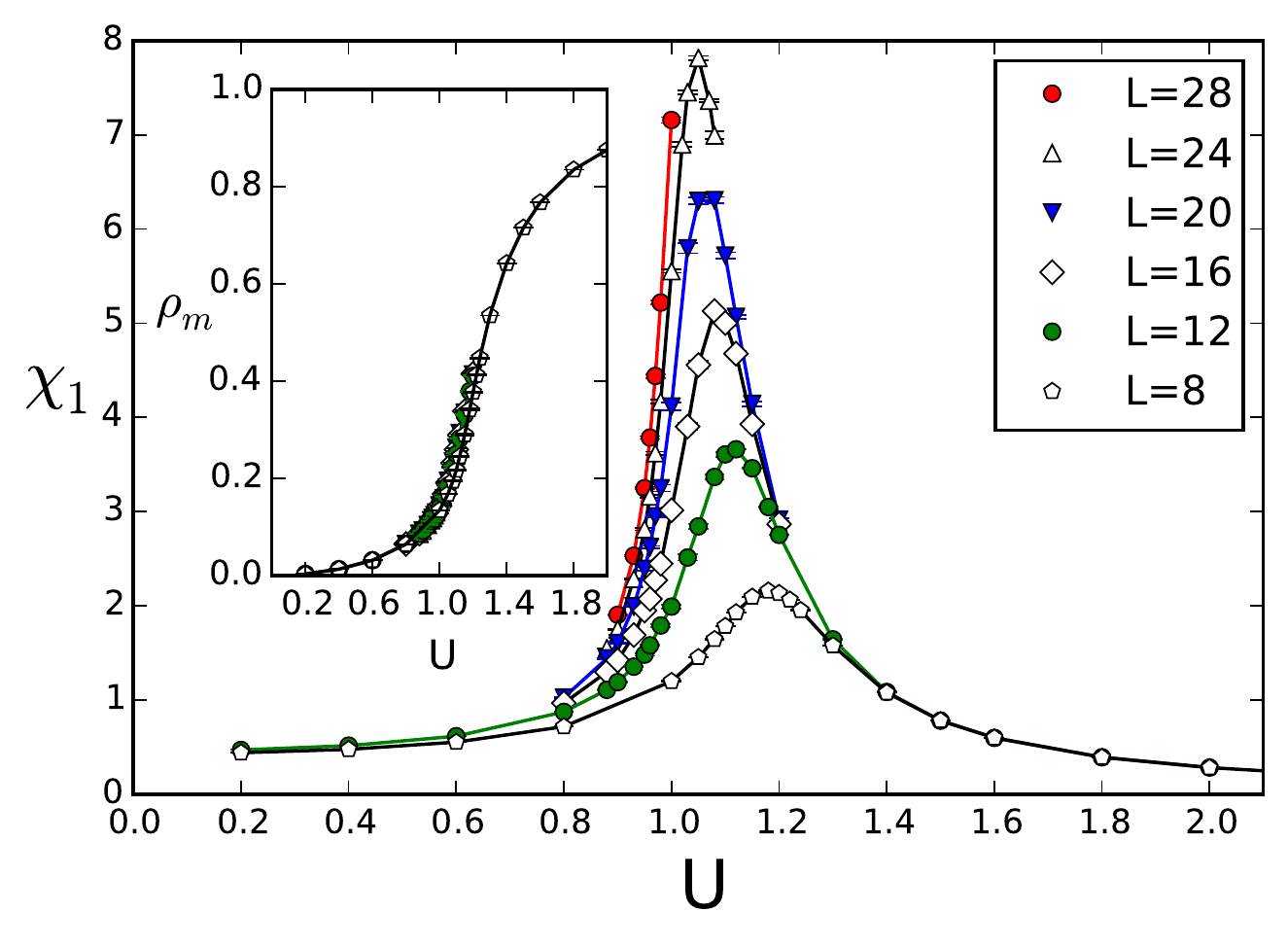}
\caption{\label{fig3} Plots of $\rho_m $ and $ \chi_1 $ as a function of $U$ for various values of $L$. The susceptibility shows a peak and the average monomer density shows a sharp rise at the phase boundary ($ U \sim 1 $).}
\end{figure}

In our earlier work we presented evidence for a single continuous phase transition between the massless and the massive phases up to lattice sizes of $L=28$. The main result is summarized in Fig.~\ref{fig3} where we plot the monomer density $\rho_m = U\langle \psi_{x,1} \psi_{x,2} \psi_{x,3} \psi_{x,4}\rangle$ and one of the fermion bilinear susceptibilities $\chi_1 = \sum_{x} \langle \psi_{0,1} \psi_{0,2} \psi_{x,1} \psi_{x,2} \rangle$, as a function of $U$ for various values of $L$. The behavior of these observables is consistent with a single phase transition around $U\approx 1$. Most importantly, the bilinear susceptibility never increases like $L^3$ showing the absence of any local fermion bilinear condensate for all values of $U$. Recently it was also confirmed that other discrete lattice symmetries, like the shift symmetry, also remain unbroken for all values of $U$ \cite{Catterall:2015zua}.

\begin{figure}[t]
\vskip0.2in
\includegraphics[width=0.48\textwidth]{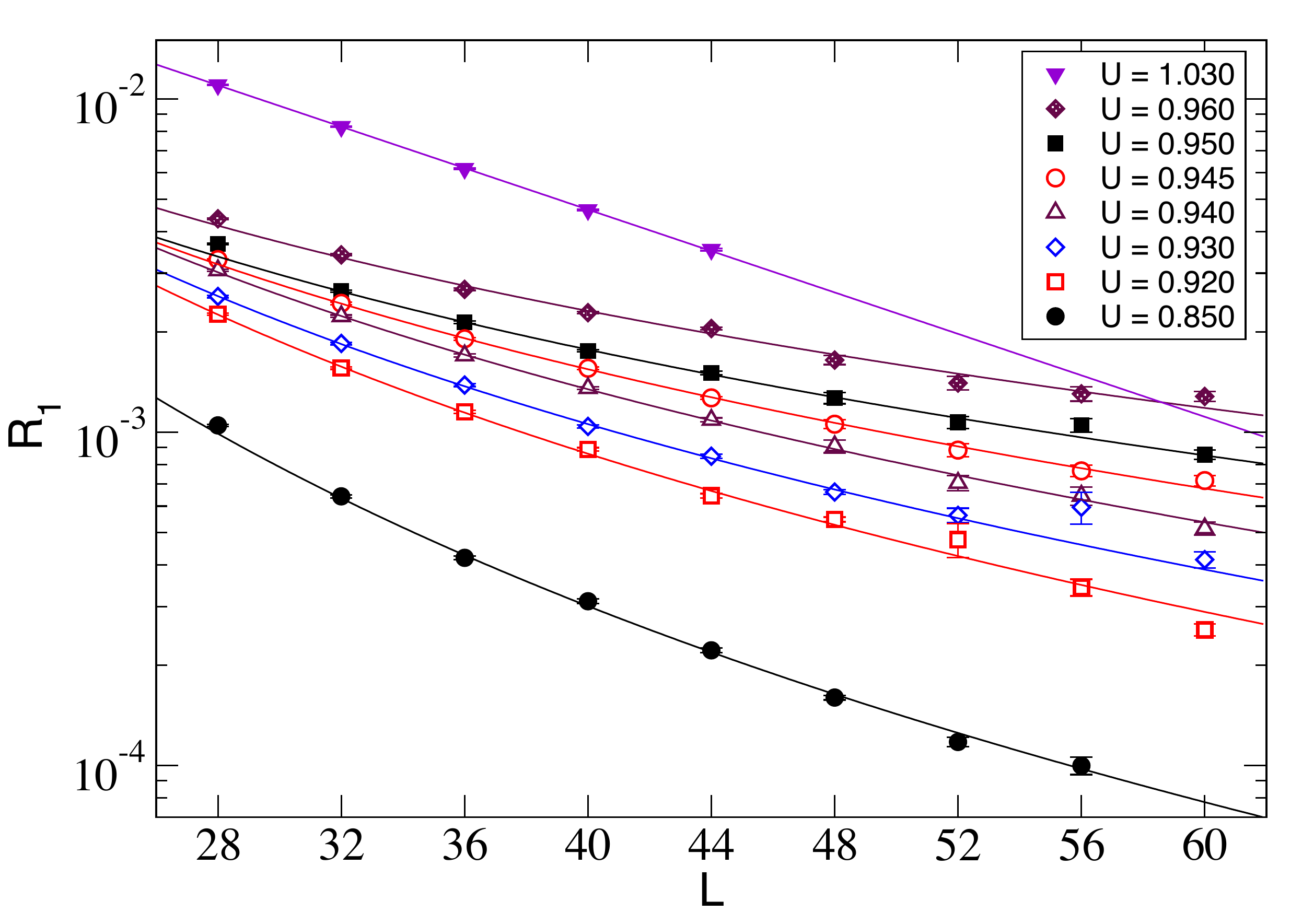}
\caption{\label{CvsL} Plot of $R_1 $ as a function of $L$ for various values of $U$ near the critical region. The solid lines are fits to the form $1/L^{1+\eta}$ where $\eta$ values are given in table \ref{eta_Uc_table}, except at $U=1.03$ where the solid line has the form $\exp(-0.07 L)$ suggesting the fermions are already massive.}
\end{figure}

We now have results from much larger lattices (up to $L=60$) that further confirm a single second order transition. We can also roughly estimate the critical exponents if we assume the absence of corrections to scaling on lattices above $L=36$. Here we focus on the two independent bosonic correlation functions $C_1(0,x) = \langle  {\psi}_{0,1}{\psi}_{0,2} \psi_{x,1} \psi_{x,2} \rangle$ and $C_2(0,x) = \langle  {\psi}_{0,1}{\psi}_{0,2} \psi_{x,3} \psi_{x,4} \rangle$ where $x$ is varied along the time direction. Near the critical point both these correlation functions are comparable to each other, while $C_2(0,x)$ vanishes at $U=0$. For the purpose of comparing different lattice sizes, we extract the correlation ratios $R_1 = C_1(0,\frac{L}{2}-1)/C_1(0,1)$ and $R_2 = C_2(0,\frac{L}{2})/ C_2(0,0)$ as a function of $L$. For large $L$, these ratios are expected to scale as $1/L^4$ in the massless phase, as $1/L^{1+\eta}$ at the critical point and as $\exp(-mL)$ in the massive phase. Here $\eta$ is one of the standard critical exponents. Our data is consistent with this behavior for $L\geq 32$. In table \ref{eta_Uc_table} we show the combined fit results of our data to the form $1/L^{1+\eta}$ near the critical region. As an illustration of the goodness of our fits, in Fig.~\ref{CvsL} we plot $R_1$ as a function of $L$ along with the fits. Based on this we estimate that the critical point is somewhere in the region $0.930 < U < 0.96$. For $U \geq 0.96$ a single power law no longer fits the data well, but an exponential fit begins to work well. For example, a fit to the form $R_1 \sim \exp(-0.07 L)$ at $U = 1.03$ is shown in Fig.~\ref{CvsL}.
\begin{table}
\begin{tabular}{|c|c|c||c|c|c|}
\hline
$U$ & $\eta$ & $ \chi^2$ & $U$ & $\eta$ & $\chi^2$ \\
& & /DOF &  & & /DOF\\
\hline
0.000 & $3$ & --  & 0.850 & $2.34(4)$  & 2.5 \\
0.920 & $1.64(5)$ & 4.6 & 0.930 & $1.44(3)$  & 1.9 \\
0.940 & $1.22(2)$ & 1.0 & 0.945 & $1.00(2) $ & 0.7 \\
0.950 & $0.77(2)$ & 1.1 & 0.960 & $0.63(5)$  & 6.4 \\
\hline
\end{tabular}
\caption{\label{eta_Uc_table} Fit results obtained by fitting both $R_1$ and $R_2$ to the form $1/L^{1+\eta}$ for various values of $U$. For small $U$ we approach $\eta \approx 3$ consistent with the free theory, while in the critical region $ 0.93 < U < 0.96 $ we again find good fits with a different $\eta$.}
\end{table}

\begin{figure}[h!]
\vskip0.2in
\includegraphics[width=0.48\textwidth]{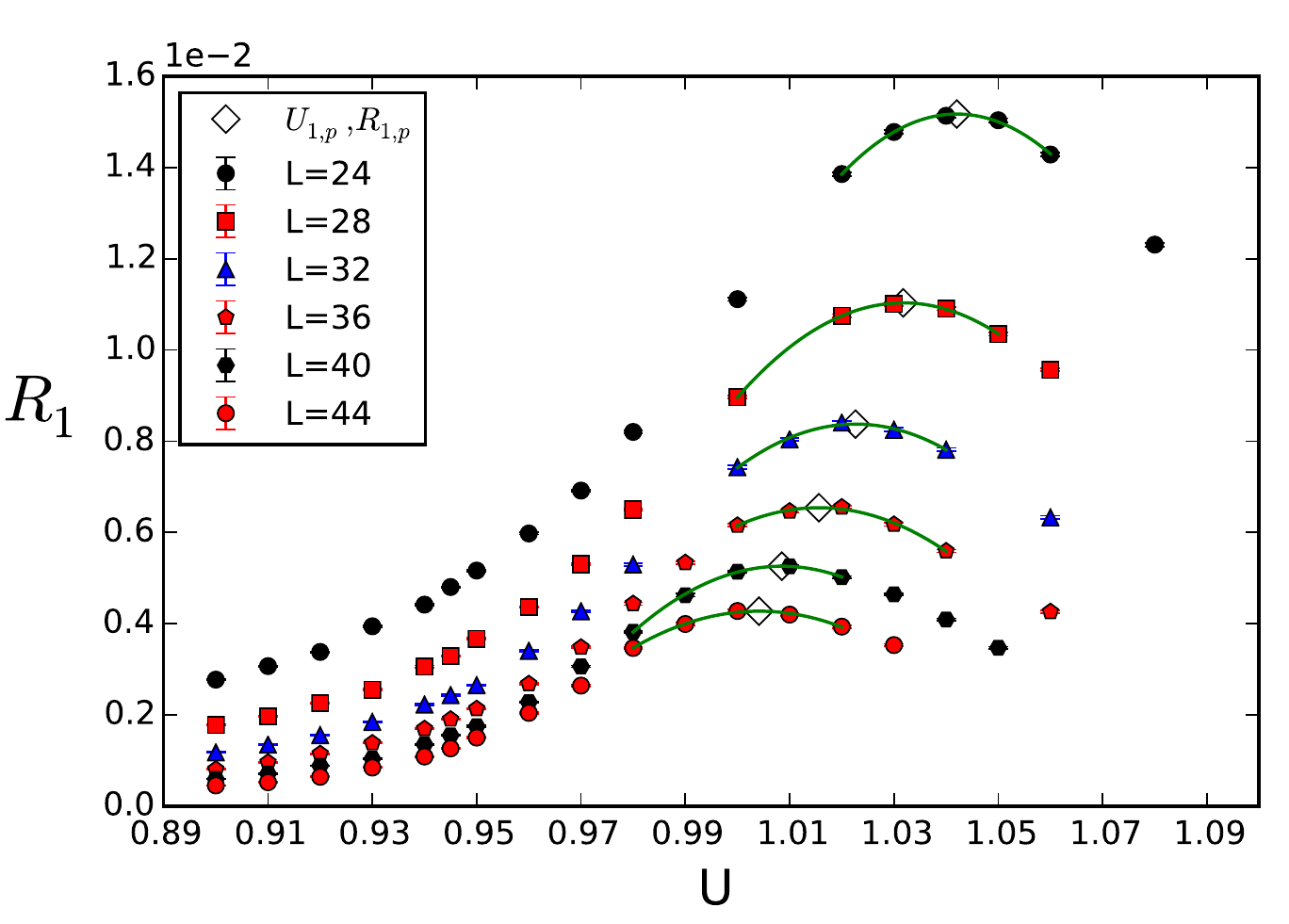}
\caption{\label{CvsU} Plots of $R_1$ as a function of $U$ for various lattice sizes showing peaks. The values of the peaks $R_{1,p}$ and their locations $U_{1,p}$ are also marked. These are determined by approximating the function to be a quadratic near the maximum.}
\end{figure}

\begin{figure}[th!]
\vskip0.2in
\includegraphics[width=0.48\textwidth]{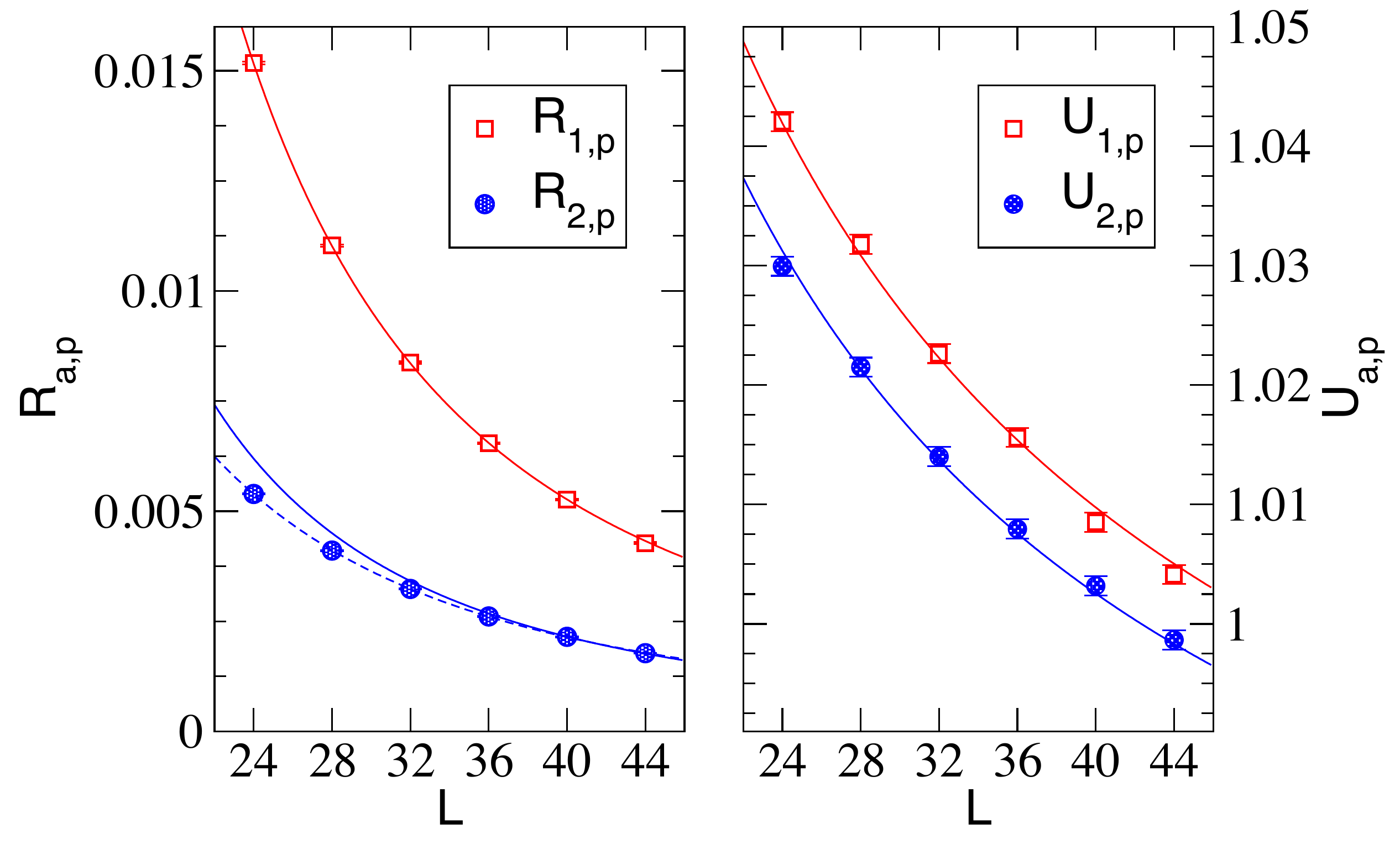}
\caption{\label{peakfit} Plots of $R_{1,p}$ and $R_{2 p}$ as a function of $L$ (left figure) and $U_{1,p}$ and $U_{2,p}$ as a function of $L$. The solid lines represent fits to the form $R_{a,p} = b_a/L^{1+\eta}$ and $U_{a,p} = U_c + d_a/L^{\nu}$ with $U_c=0.943$ fixed. The dashed line is a fit including correction to scaling of the form $R_{2,p} = b_2/L^{1+\eta}+ c_2/L^{1+\eta+\omega}$, where $\omega \approx 1$.}
\end{figure}

\begin{figure}[th!]
\vskip0.2in
\includegraphics[width=0.48\textwidth]{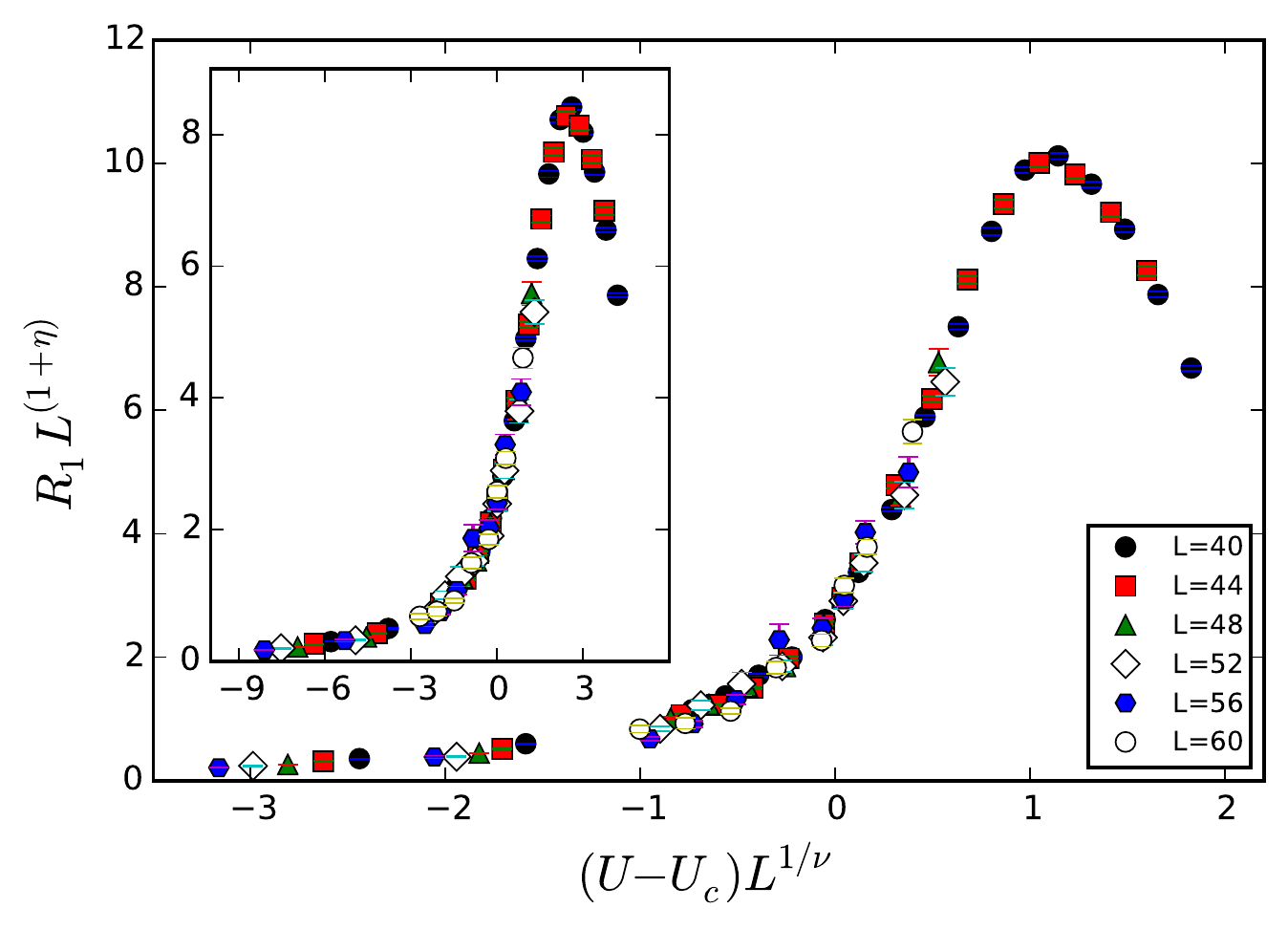}
\caption{\label{univfn} Evidence for universal scaling in our large lattice data with $U_c=0.943$, $\eta =1.08$, and $\nu = 1.30$. Our data may also be consistent with $U_c=0.945$, $\eta =1$, and $\nu = 1$ expected from large $N$ analysis (shown in the inset), but only after including corrections to scaling.}
\end{figure}

Since it is difficult to locate $U_c$ and compute $\eta$ from the data in the region $0.930 < U < 0.960$ we have also analyzed a different scaling region of $U$ where $R_1$, $R_2$ show a peak. In Fig.~\ref{CvsU} we plot the behavior of the correlation ratio $R_1$ as a function of the coupling $U$ for different lattices sizes. A similar plot exists for $R_2$ \cite{SupplMat}. Note that the correlation ratios display a maximum as a function of $U$ for a fixed $L$. We have computed these maximum values $R_{1,p}(L)$, $R_{2,p}(L)$ and their locations $U_{1,p}(L)$, $U_{2,p}$ in the range $24 \le L \le 44$. From scaling theory, we expect $R_{a,p} = b_a/L^{1+\eta}$ and $U_{a,p} = U_c + d_{a}/L^{\nu}$. We find that $R_{1,p}$ fits well to this expected form for $24 \le L \le 44$, while $R_{2,p}$ does not. However if we keep only the data from the largest lattices for both $R_{1,p}$ and $R_{2,p}$ we can again perform combined fits to the expected scaling form without the need for corrections to scaling. Interestingly, allowing a scaling correction only for $R_{2,p}$ allows us to fit the entire data set. Two of these fits are shown in the left plot of Fig. \ref{peakfit}. Using these fits and including various systematic errors we estimate $\eta= 1.05(5)$. Combining this result with that of Table \ref{eta_Uc_table}, we constrain $U_c = 0.943(2)$. Using this result along with our data for $U_{a,p}$ and its expected scaling form we can again perform combined fits to obtain $\nu$. One such fit is shown in the right plot of Fig.~\ref{peakfit}. Using these fits we estimate $\nu = 1.30(7)$. In Fig.~\ref{univfn} we verify if our large lattice data falls on a single universal scaling function when we fix $U_c=0.943$, $\eta=1.05$ and $\nu = 1.30$.  The fact that the data falls on a single curve gives us confidence that this is indeed the case. However, we must note that if we allow for scaling corrections to be present in our fits we cannot rule out $U_c=0.945$, $\eta=1.0$ and $\nu=1.0$ as expected from large $N$ four-fermion models \cite{Hands:1992be}. The universal scaling with these ``mean field'' parameters is shown in the inset of Fig.~\ref{univfn}.

To summarize, we have established the existence of a three dimensional exotic second order phase transition between a massless and a massive fermion phase, both of which have the same lattice symmetries. Thus, fermion mass generation is not necessarily driven by spontaneous symmetry breaking contrary to conventional wisdom. 

The material presented here is based upon work supported by the U.S. Department of Energy, Office of Science, Nuclear Physics program under Award Number DE-FG02-05ER41368. An important part of the computations performed in this research was done using resources provided by the Open Science Grid, which is supported by the National Science Foundation and the U.S. Department of Energy's Office of Science \cite{Pordes2008,Sfiligoi2009}.

\bibliography{ref,pms,topins}


\clearpage

\newpage

\pagestyle{empty}

\begin{center}
{\Large Supplementary Material}
\end{center}
In this supplementary material we explain the analysis used to arrive at the results presented in the main paper.
\section*{Observables}
The three quantities we focus on in the paper are the monomer density and the two independent correlation functions defined as
\begin{eqnarray}
\rho_m &=&  U\langle \psi_{x,1} \psi_{x,2} \psi_{x,3} \psi_{x,4}\rangle, \\
C_1(x) &=&  \langle  {\psi}_{0,1}{\psi}_{0,2} \psi_{x,1} \psi_{x,2} \rangle, \\
C_2(x) &=&  \langle  {\psi}_{0,1}{\psi}_{0,2} \psi_{x,3} \psi_{x,4} \rangle.
\end{eqnarray}
The correlation functions are defined between the origin and an arbitrary lattice site $x$ that varies along the time direction.

In principle there are six onsite fermion bilinear operators through which we can define $36$ different correlation functions. However, due to the $SU(4)$ symmetry of the model all these correlation functions are related to the above two independent correlation functions. If we consider the origin as an even site, it is easy to argue that for a non-vanishing correlation function, $x$ must be an odd site in the case of $C_1(x)$ and it must be an even site in the case of $C_2(x)$. The monomer density is related to the second correlation function through the relation $\rho_m \ = \ U\ C_2(0)$. Since $\rho_m \sim U^2$ for small $U$, we can argue that $C_2(x)$ approaches zero at small $U$ but not $C_1(x)$. On the other hand close to the critical point the two correlation functions are very similar and are expected to scale identically with respect to $x$ for large values of $x$. In the earlier work we measured the susceptibilty
\begin{equation}
\chi_a = \sum_x C_a(x),
\end{equation}
which turned out to be computationally expensive with our modified algorithm that allows us to explore large lattices. Hence we have focused on the correlation ratios in this work, which are defined as
\begin{eqnarray}
R_1 &=& C_1(0,\frac{L}{2}-1)/C_1(0,1),\\
R_2 &=& C_2(0,\frac{L}{2})/ C_2(0,0).
\end{eqnarray}
These ratios help us extract the interesting infrared physics, by removing any multiplicative renormalizations that may arise from the ultraviolet. 

\section{Critical Scaling}
The correlation ratios $R_1$ and $R_2$ depend on both $U$ (the coupling) and $L$ (the lattice size) and near the critical point they are expected to scale according to the form :
\begin{eqnarray}
R_a(U,L)&=& \frac{1}{L^{(1+\eta)}} \ g_a\left( (U-U_c) \ L^{\frac{1}{\nu}} \right) \label{scal} 
\end{eqnarray}
where $g_a(x),\ a=1,2$ are universal functions of the variable $x = (U-U_c) \ L^{\frac{1}{\nu}}$. We use this behavior to extract the critical exponents in our work. For example when $U=U_c$ we expect
\begin{equation}
R_a(U_c,L) = \frac{f_a}{L^{1+\eta}}
\label{etacrit}
\end{equation}
where $f_a$ is a constant and the critical exponent $\eta$ is the same for both $R_1$ and $R_2$. Interestingly even in the free theory $U=0$ one expects the above form to hold except that $\eta=3$. In fact for sufficiently large $L$, we expect $\eta=3$ in the entire massless phase.

\begin{figure}[htb!]
\includegraphics[width=0.45\textwidth]{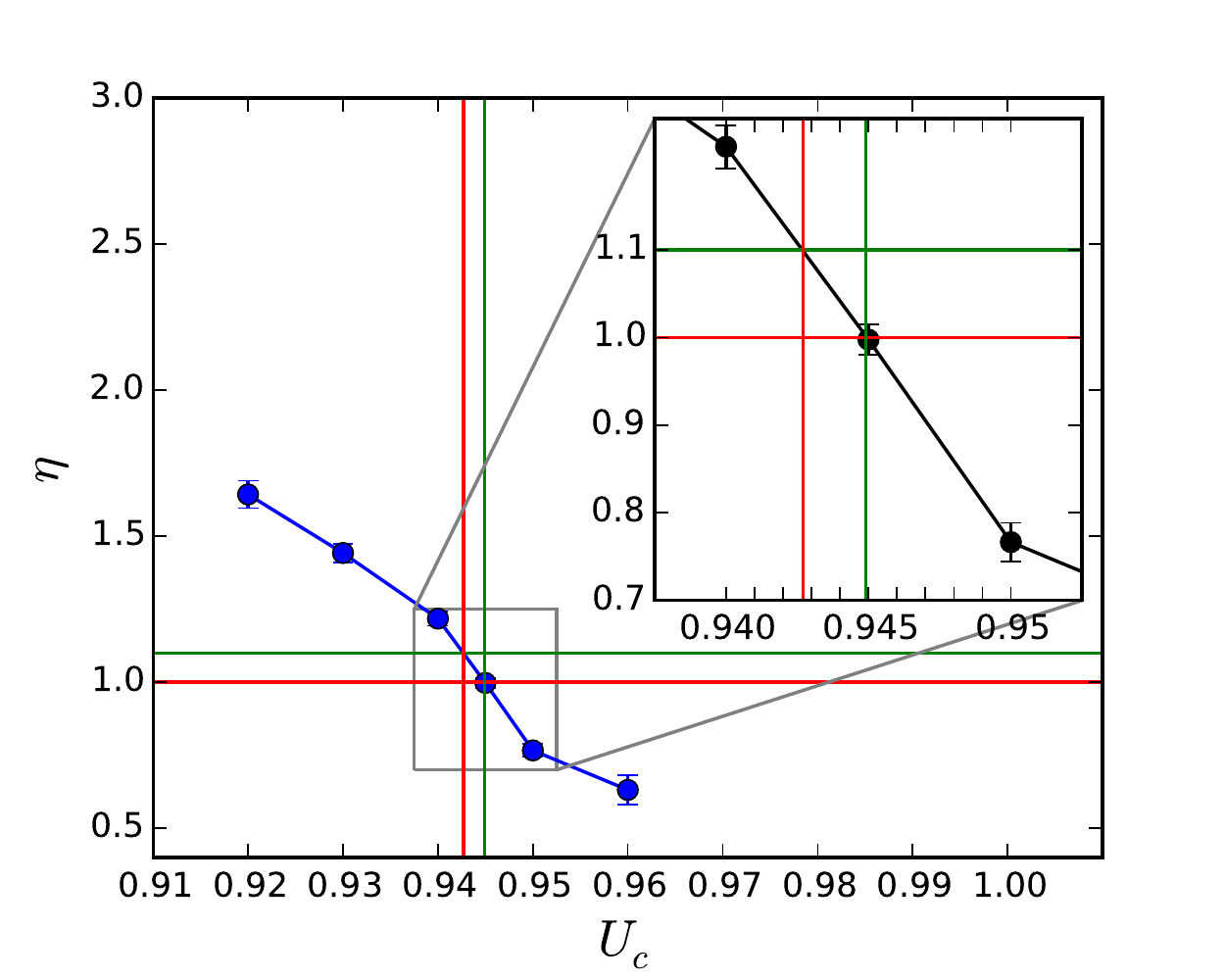}
\caption{\label{eta_Uc} Plot of $ \eta $ as a function of $ U_c $. We have used the power-law fits to obtain the $\eta$ values assuming the location of the critical point.  }
\end{figure}

Unfortunately since we do not know the location of the critical point nor the value of $\eta$ at that point, it is difficult to compute $U_c$ and $\eta$ together using the above relation. This is typical of all second order critical points and one usually finds that many couplings near the critical point obey power law scaling with slightly different values of $\eta$. A combined fit of both correlation ratios to the form given in (\ref{etacrit}) for the couplings in the range $0.85 \leq U \leq 0.96$ and lattice sizes $L \geq 32$, yields the results shown in Table \ref{eta_Uc_table_2}. 
\begin{table}[htb!]
\begin{tabular}{|c|c|c|c|c|}
\hline
$ U_c $ & $f_1$ & $f_2$ & $\eta$ & $ \chi^2 $ \\
\hline
0.85	& 68(10) & 38(5)  & $2.34(4)$	& 2.2	 \\ 
0.92	& 15(3) & 8(1)  & $1.64(5)$	& 4.1	 \\ 
0.93	& 9(1) & 4.5(5)  & $1.44(3)$	& 1.9	 \\
0.94	& 4.8(4) & 2.4(2)  & $1.22(2)$	& 1.0	 \\
0.945 & 2.5(2) & 1.2(1) & $1.00(2) $	& 0.7	 \\
0.95	& 1.2(1) & 0.59(5) & $0.77(2)$	& 1.1	 \\
0.96	& 1.0(2) & 0.46(8) & $0.63(5)$	& 6.4	 \\
\hline
\end{tabular}
\caption{\label{eta_Uc_table_2} Results for the critical exponent $\eta $ and the critical coupling $U_c$ from the powerlaw fits. }
\end{table}
Indeed the critical point could be at any value of $U$ in the range $0.93 < U_c < 0.96$. The behavior of $\eta$ as a function of $U$  is shown in Fig \ref{eta_Uc}. In our analysis we will use an independent procedure to estimate $\eta$ and then use it to constrain $U_c$ along with this figure. Note that for $U=0.96$ the fit is poor, indicating perhaps that we have reached the massive phase. 

\section{Scaling of Pseudo-Critical Points}

\begin{figure*}[htb!]
\hbox{
\includegraphics[width=0.48\textwidth]{fig5.pdf}
\includegraphics[width=0.48\textwidth]{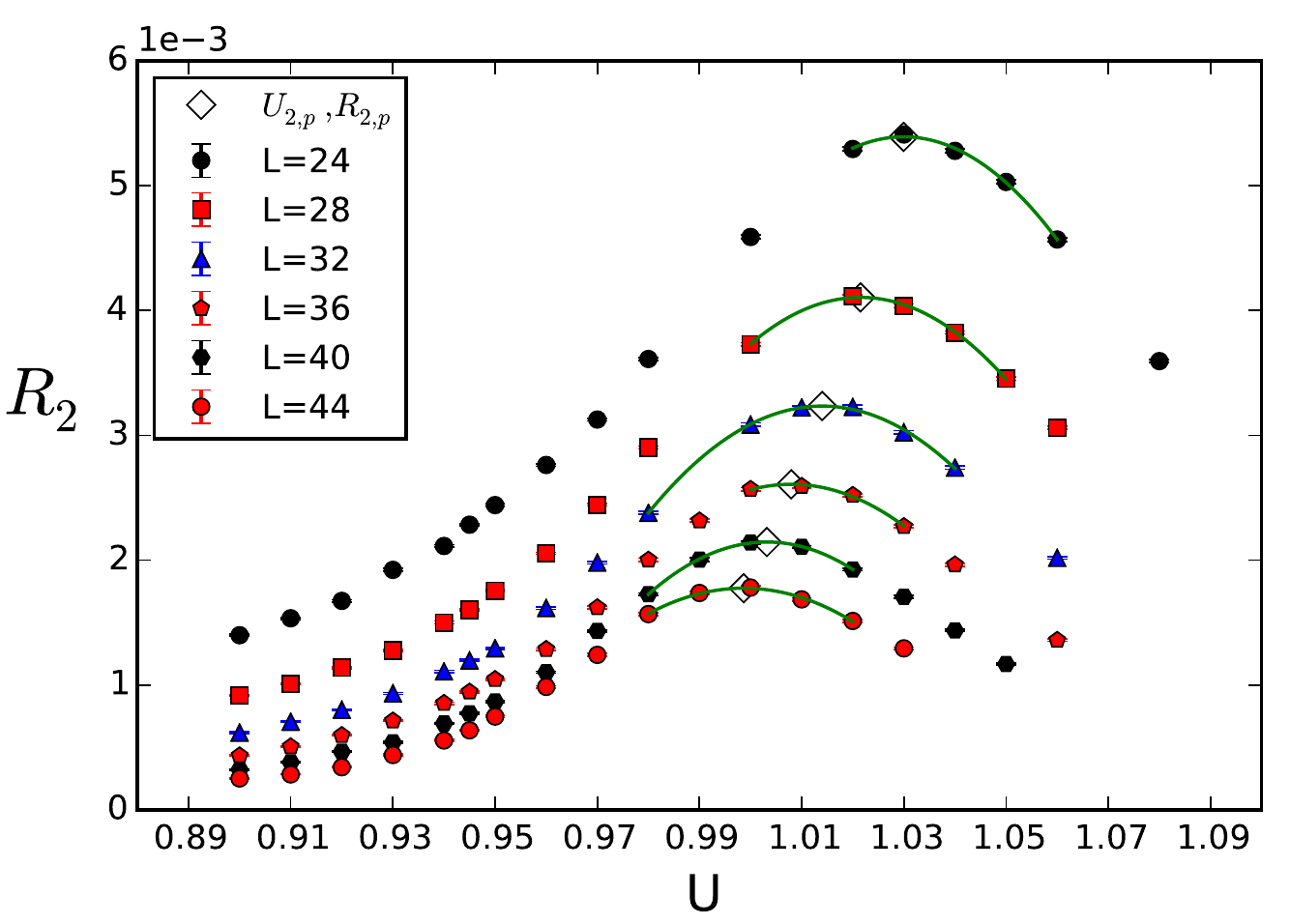}
}
\caption{\label{C_v_U} Plot of correlation ratios $R_a, a = 1,2$ as a function of coupling $U$. These ratios display a maximum at $U_{a,p}$.}
\end{figure*}

While critical points are defined only in the thermodynamic limit, there exists a notion of pseudo-critical points even in a finite system. Interestingly, quantities close to the pseudo-critical points also obey critical scaling and thus can help in the extraction of critical quantities independently. Consider for example the variation of correlation ratios $R_1$ and $R_2$ with coupling $U$ at a fixed value of $L$. This is shown in Fig. \ref{C_v_U} for different lattice sizes. It is clear from the figure that the ratios display a maximum for certain value of the coupling (which we refer to as $U_{a,p}$). These define pseudo-critical couplings. At the peak, the value of the ratio itself is given by $R_{a,p}$. From the scaling relation in eqn \ref{scal}, we note that the peak occurs when the function $g_a(x)$ reaches a maximum. Assuming this occurs at $x= d_a$, we can derive
\begin{eqnarray}
R_{a,p} &=& \frac{b_a}{L^{1+\eta}} \label{etaq}\\
U_{a,p} &=& U_c +  d_a/L^{\frac{1}{\nu}} \label{nup}.
\end{eqnarray}
Thus, if we can compute $R_{a,p}$ as a function of $L$, we would have an independent way to estimate $\eta$ using (\ref{etaq}). As we explained above, we can then use the $\eta$ vs. $U_c$ plot of Fig.~(\ref{eta_Uc}), in order to estimate $U_c$. Then using this value of $U_c$ in (\ref{nup}) we can compute $\nu$. This will be our strategy.

We approximate the behavior of the correlation ratios around the peak as a quadratic to extract $U_{a,p}$ and $R_{a,p}$. Table \ref{table_fits2} shows our results of such fits.  The errors in the fits shown include systematic errors associated with choosing a quadratic form near the peak instead of say a cubic or quartic form. 
\begin{table*}[htb!]
\begin{tabular}{|c|c|c|c||c|c|c|}
\hline
$ L $ 	 & $ U_{1,p} $ 	 & $ R_{1,p} $ 	 & $ \chi^2$/dof &
$ U_{2,p} $ 	 & $ R_{2,p} $ 	 & $ \chi^2$/dof \\
\hline
\hline
24.0	& $1.0420(8)$	& $1.517(3) \times 10^{-02} $	& 0.473 &	
 $1.0299(8)$	& $5.391(9) \times 10^{-03} $	& 1.734	 \\
28.0	& $1.0318(8)$	& $1.103(3) \times 10^{-02} $	& 0.1622	 &
 $1.0215(8)$	& $4.105(9) \times 10^{-03} $	& 0.8016	 \\
32.0	& $1.0226(8)$	& $8.38(3) \times 10^{-03} $	& 1.519	 &
 $1.0140(8)$	& $3.235(9) \times 10^{-03} $	& 1.407	 \\
36.0	& $1.0156(8)$	& $6.54(3) \times 10^{-03} $	& 1.752	 &
 $1.0080(8)$	& $2.608(9) \times 10^{-03} $	& 2.004	 \\
40.0	& $1.0085(8)$	& $5.26(3) \times 10^{-03} $	& 0.4788	 &
 $1.0032(8)$	& $2.146(9) \times 10^{-03} $	& 0.21	 \\
44.0	& $1.0041(8)$	& $4.28(3) \times 10^{-03} $	& 0.7981	 &
 $0.9986(8)$	& $1.776(9) \times 10^{-03} $	& 0.9341	 \\
\hline
\end{tabular}
\caption{\label{table_fits2} Results for the value of $R_{1,p}$, $U_{1,p}$, $R_{2,p}$ and $U_{2,p}$ obtained from a quadratic fit of the data near the peak.}
\end{table*}

\subsection{Scaling Fits for $R_{a,p}$}
\begin{figure*}[htb!]
\includegraphics[width=0.48\textwidth]{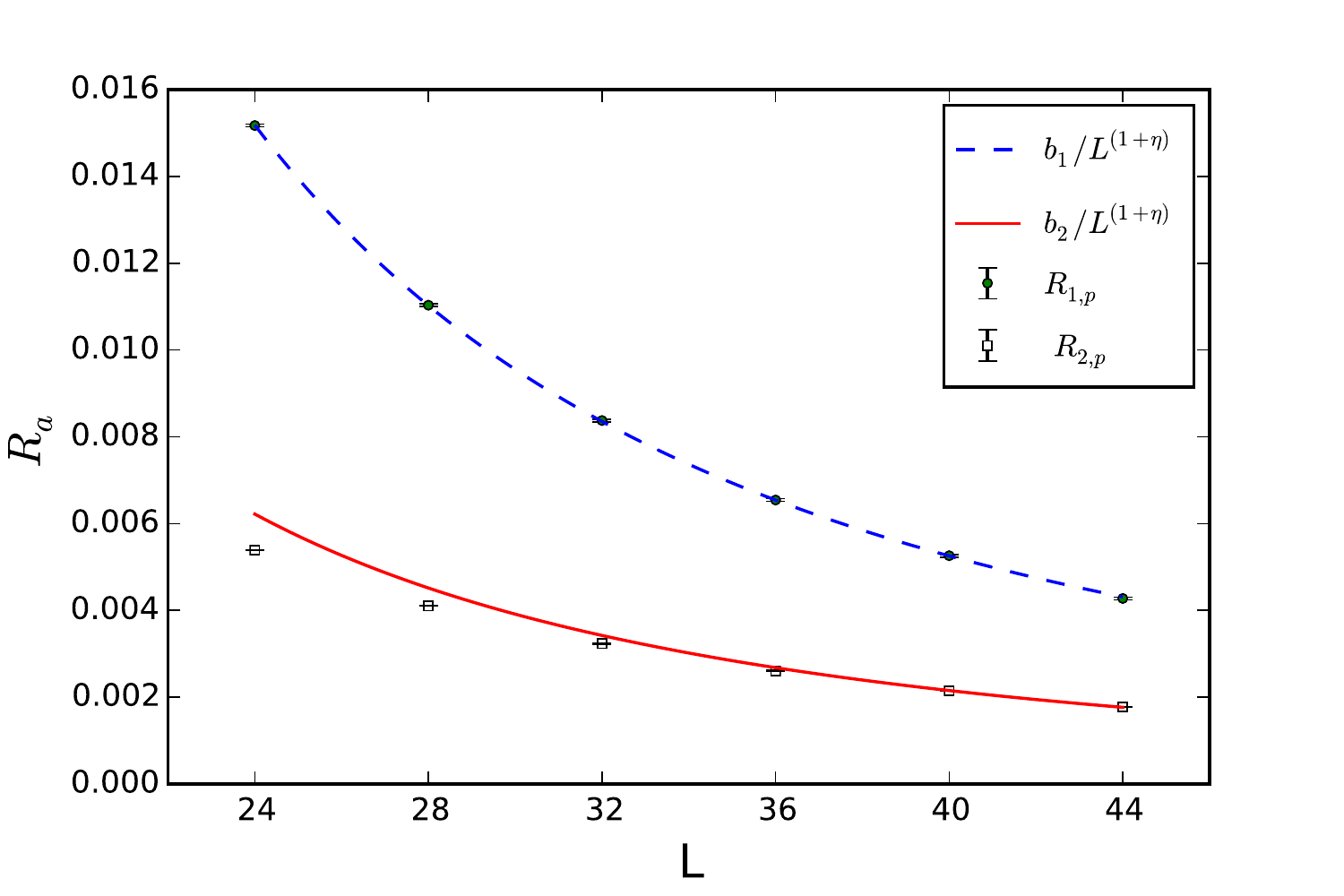}
\includegraphics[width=0.48\textwidth]{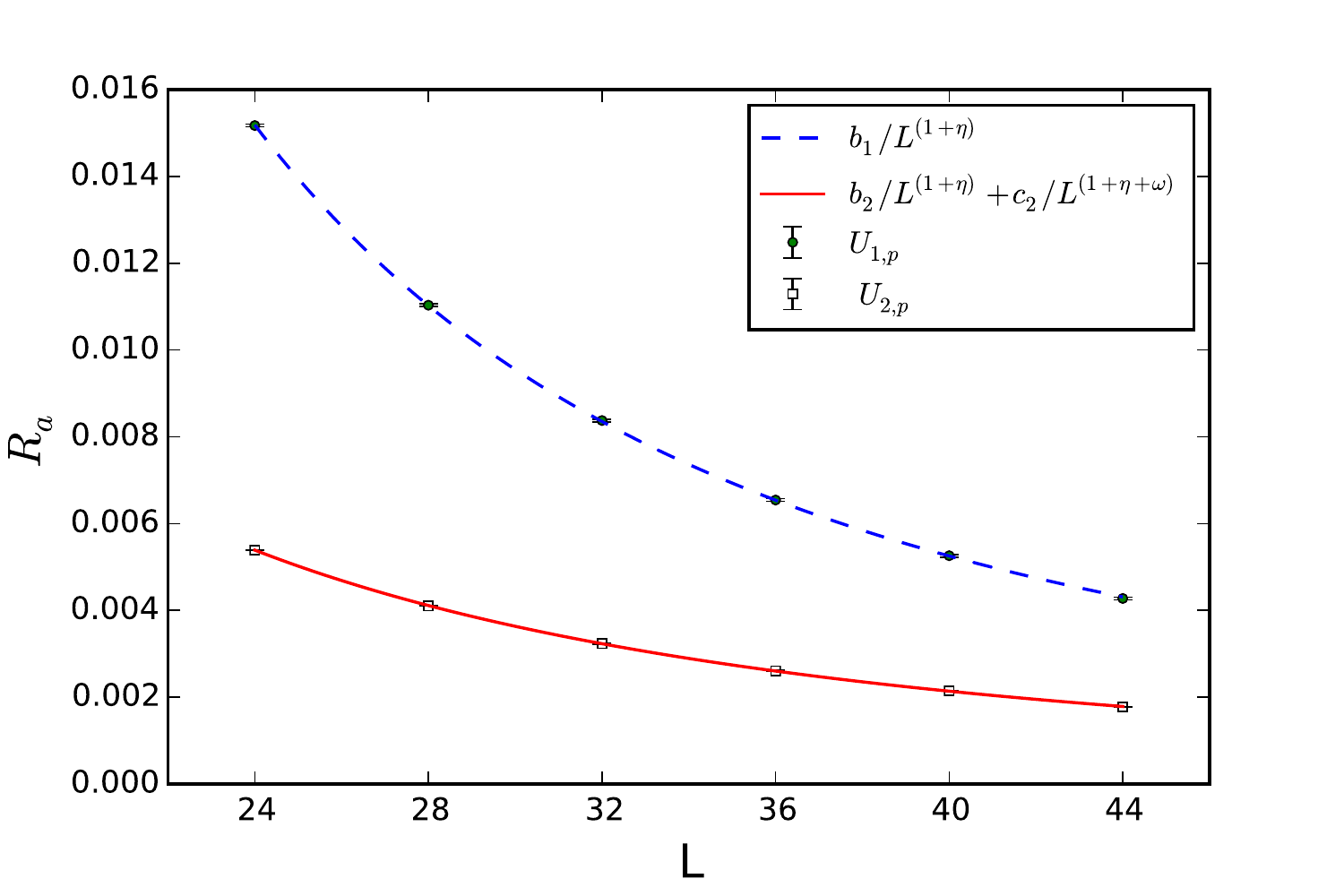}
\caption{\label{Rpfits} Plots of $R_{1,p}$ and $R_{2,p}$ as a function of $L$. On the left, the solid and dashed lines show fits to the form (\ref{etaq}), while on the right these lines show fits to the form (\ref{etapc1}) and (\ref{etapc2}) that includes correction to scaling. To achieve a good fit, in the left figure it was necessary to drop the data for $R_{2,p}$ in the region $L\leq 36$, while in the right figure we could fit the entire data by assuming the correction to the scaling exponent $\omega =1$.}
\end{figure*}

Using the data for $R_{1,p}$ and $R_{2,p}$ from table \ref{table_fits2} we have performed a combined fit of the form expected in (\ref{etaq}). Including the entire data from above ($ 24 \le L \le 44 $) gives us $b_1 =6.3(9)$, $b_2 =2.4(4)$, $\eta=0.91(4)$,  $\chi^2= 94.9$. A closer examination shows that while $R_{1,p}$ fits well to a single power law in the entire region giving $b_1 = 11.2(2)$, $\eta = 1.08(1)$ with a $\chi^2/d.o.f = 1$, $R_{2,p}$ is not consistent with a single power law. Table \ref{tabR2} shows the results of fitting $R_{2,p}$ individually and dropping the lower lattice sizes systematically.
\begin{table}[htb!]
\begin{tabular}{|c|c|c||c|}
\hline
$L$-Range & $b_2$ & $\eta$ & $ \chi^2$/dof \\
\hline
24-44	& $1.68(3)$	& $0.81(1)$	& 6.3 \\
28-44	& $1.84(5)$	& $0.83(1)$	& 3.9 \\
32-44	& $2.10(11)$	& $0.86(2)$	& 2.3 \\
36-44	& $2.38(25)$	& $0.90(3)$	& 2.2 \\
40-44	& $3.25(76)$	& $0.99(6)$	& 0.0 \\
\hline
\end{tabular}
\caption{\label{tabR2} Fits of $R_{2,p}$ as a function of $L$ to the expected scaling form for different ranges of lattice sizes. Importantly $\eta$ drifts upwards.}
\end{table}
We interpret the drift of $\eta$ to larger values as a sign that $R_{2,p}$ contains pronounced corrections to scaling. If we only keep the lattice sizes of $L=40,44$ in the $R_{2,p}$ data and perform a combined fit of both $R_{1,p}$ and $R_{2,p}$ we obtain $b_1=11.1(2)$, $b_2=4.6(1)$, $\eta=1.08(1)$, $\chi^2=0.77$. The goodness of the fit is shown in the left plot of Fig.~\ref{Rpfits}.

In order to confirm that the drift of $\eta$ is consistent with the presence of corrections to scaling, we added a correction term for $R_{2,p}$ and performed a combined fit of the entire data to the form:
\begin{eqnarray}
 R_{1,p}&=& \frac{b_1}{L^{1+\eta}} \label{etapc1} \\
 R_{2,p}&=& \frac{b_2}{L^{1+\eta}} + \frac{c_2}{L^{1 +\eta + \omega }}
\label{etapc2} 
\end{eqnarray}
Now including the entire data set in table \ref{table_fits2}, we find a good fit as shown in the right plot of Fig.~\ref{Rpfits} , giving us $b_1=11.2(2)$, $b_2 = 5.6(4)$, $c_2=-25(12)$, $\eta= 1.08(1)$,  $\omega= 0.9(2)$, $\chi^2= 0.752$. This gives some credence to our belief that $R_{2,p}$ data contains corrections to scaling. 

However, there is a bias in the above analysis since it is likely that the presence of smaller lattice data in $R_{1,p}$ affects the fitting. Hence we roughly estimate the systematic errors in $\eta$ due to the the range of lattice sizes we use in the fit. Keping only $L=40,44$ data for both $R_{1,p}$ and $R_{2,p}$ and ignoring corrections to scaling we obtained $\eta \approx 1.05$ but with a $\chi^2/dof = 2.5$ which is rather large. But by keeping $L=36$ and dropping $L=44$ instead we get a good fit but with $\eta \approx 1.02$. Thus, a conservative estimate would be $\eta = 1.05(5)$. 

\begin{figure}[htb!]
\vskip0.2in
\includegraphics[width=0.45\textwidth]{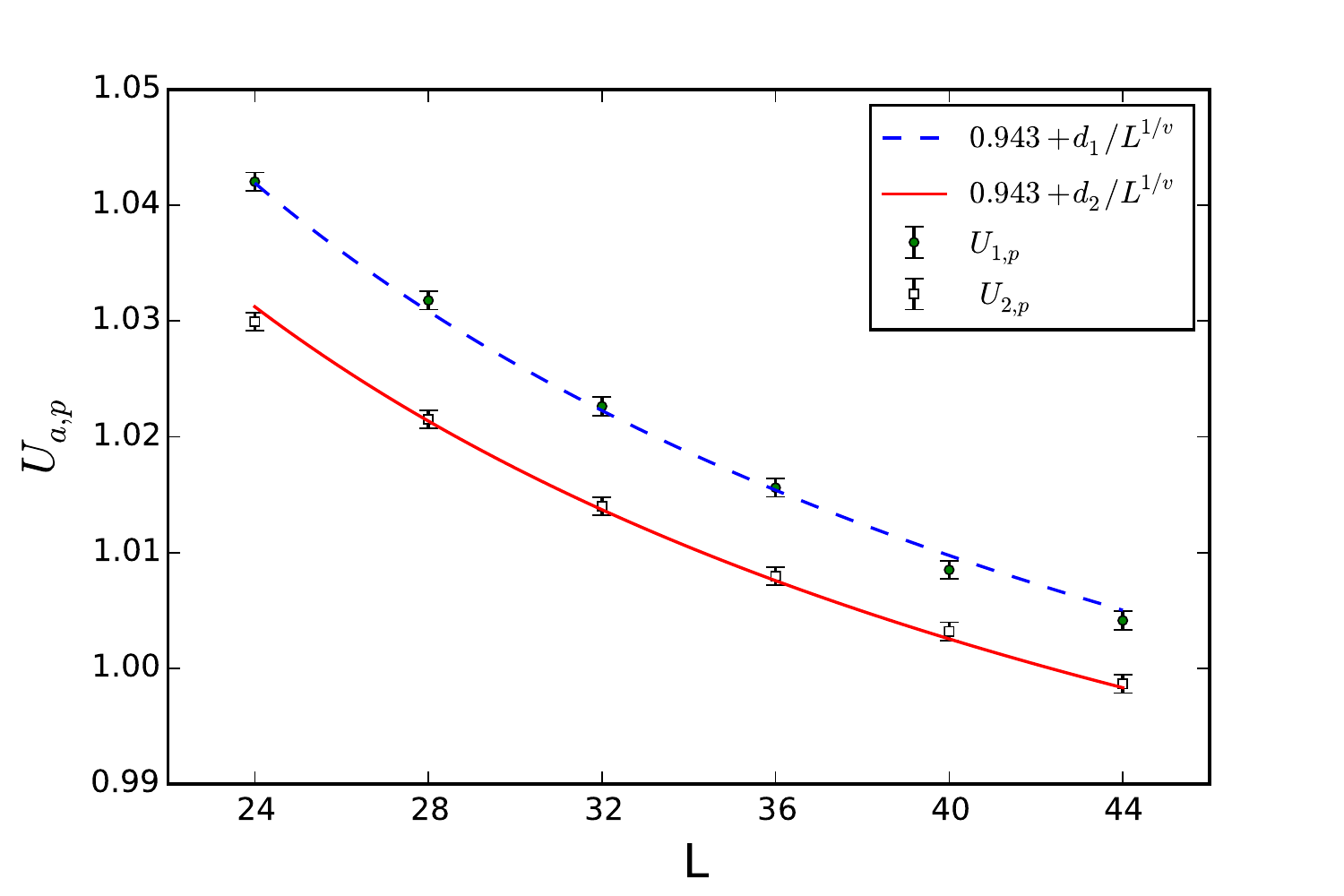}
\caption{\label{nu_fit} Plots of $U_{1,p}$ and $U_{2,p}$ as a function of $L$. The solid and dashed lines are fits to (\ref{nup}) assuming $U_c = 0.943$.  }
\end{figure}

\subsection{Scaling Fits for $U_{a,p}$}

Using $\eta = 1.05(5)$ we can again conservatively estimate $U_c$ from Fig.~\ref{eta_Uc} to be $ U_c = 0.943(2)$. We can then use (\ref{nup}) to estimate $\nu$. Again assuming no corrections to scaling we find that a combined fit of both the data $U_{1,p}$ and $U_{2,p}$ fits well to single power law. Performing two fits by fixing $U_c=0.945$ and $U_c=0.941$ we obtain $\nu=1.30(7)$. The goodness of the fits, assuming $U_c=0.943$, are shown in Fig.~(\ref{nu_fit}). For this value of $U_c$ we obtain $d_1 = 1.14(6)$ and $d_2=1.02(5)$ and the $\chi^2$/dof = 1.0.

\section{Corrections to Scaling}

Since we have ignored corrections to scaling in the analysis above, one might wonder if introducing corrections to scaling can change the results. Experience tells us that usually once we include corrections to scaling the fits become unstable unless we can constrain at least some of the exponents by other arguments. This is difficult in our context due to the exotic nature of the phase transition. There is little information to go by. Still we can ask if, for example, our data is consistent with the exponents from large $N$ predictions in a typical Gross Neveu model, i.e., $\eta=1$ and $\nu=1$.

\begin{figure*}[htb!]
\includegraphics[width=0.45\textwidth]{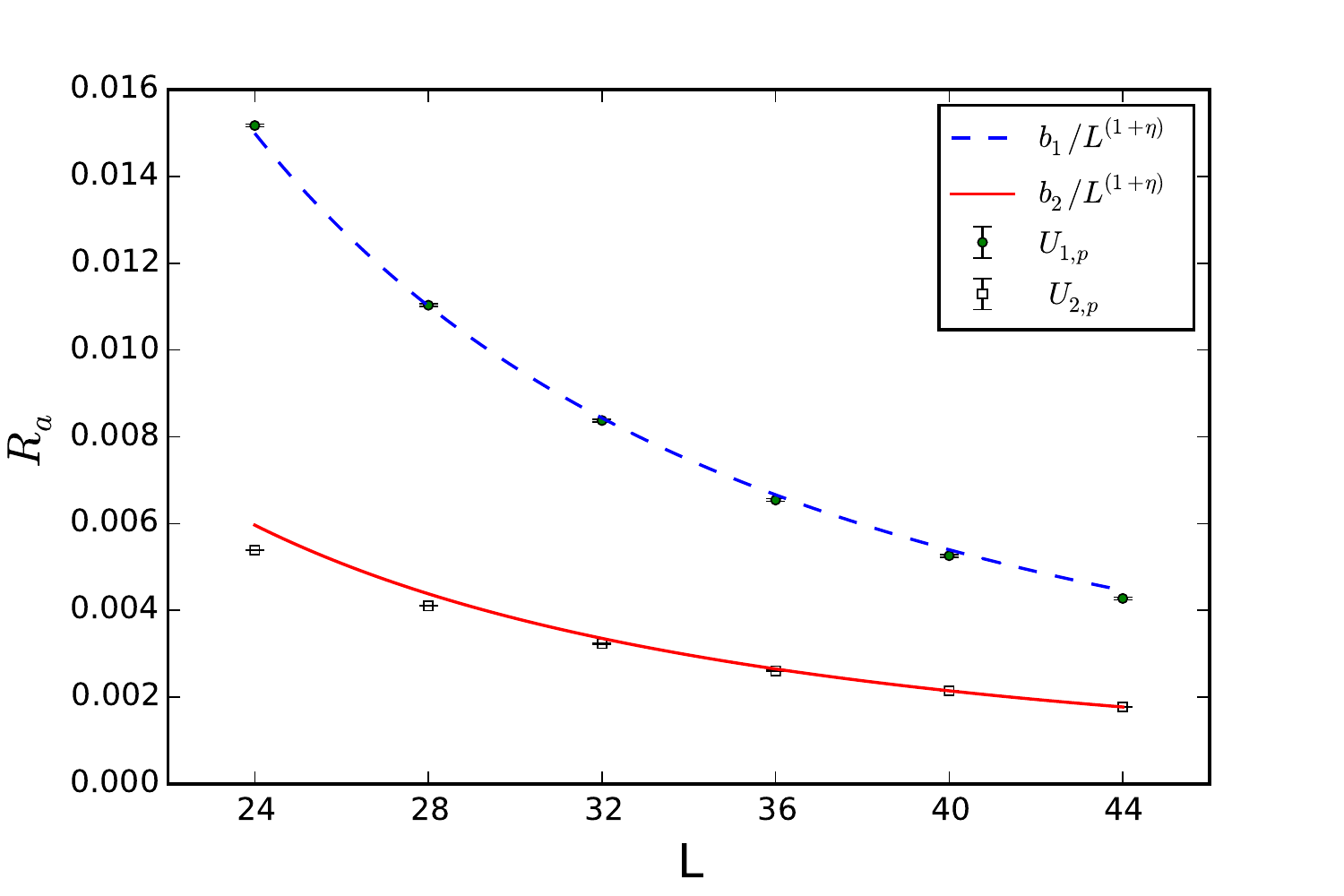}
\includegraphics[width=0.45\textwidth]{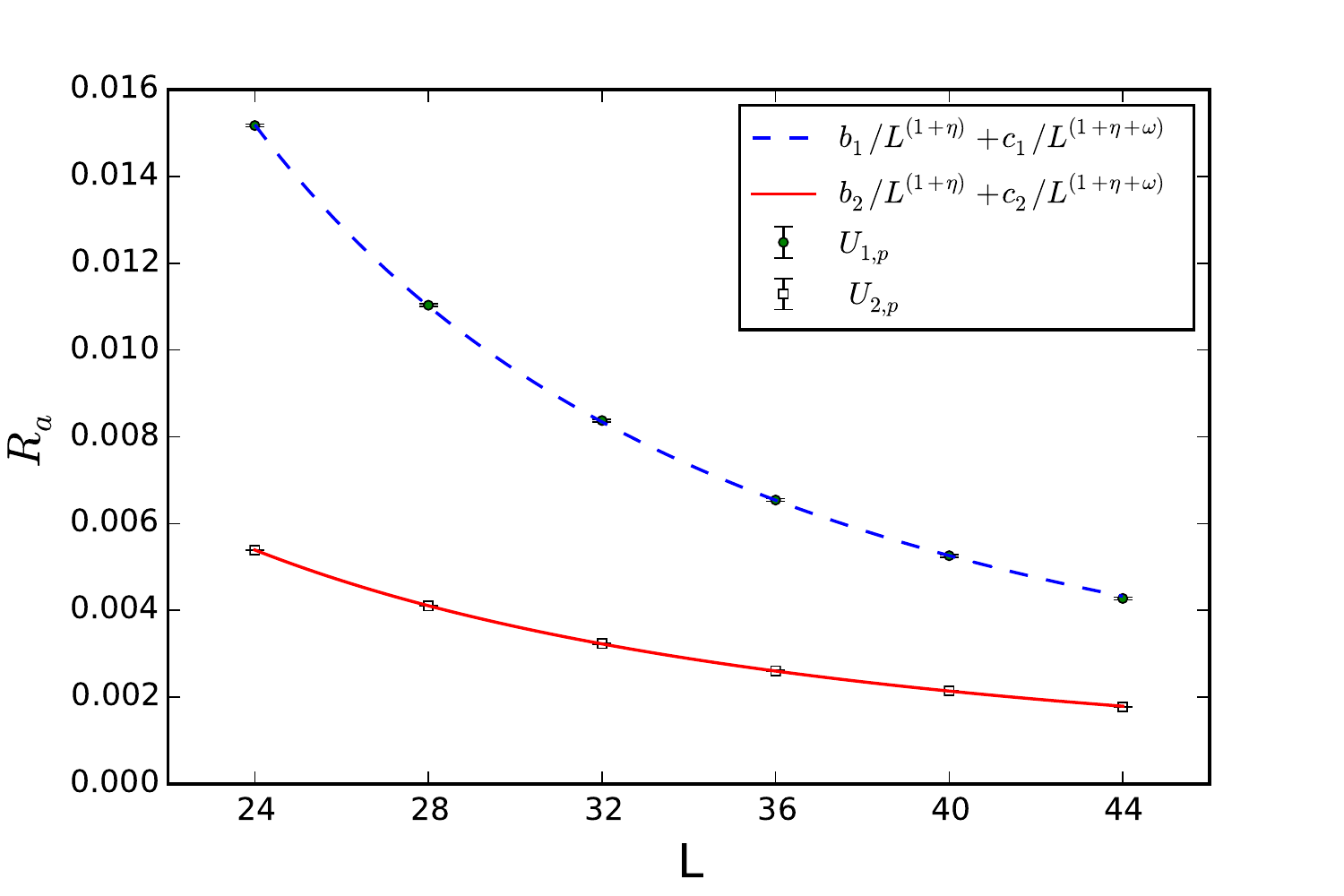}
\caption{\label{lnfits} Plots of $R_{1,p}$, $R_{2,p}$ as a function of $L$ showing fits to large $N$ exponents with (right) and without (left) corrections to scaling. In the fits we fix $\eta=1$ and $\omega=1$. The $\chi^2$/dof for the fits are $21$ when the corrections to scaling are omitted and $1.1$ when they are included.} 
\end{figure*}

Assuming no corrections to scaling, but fixing $\eta=1$ and removing data for $L =24,28,32,36$ for $ R_{2,p}$ which gave a good fit above yields $b_1 =8.5(1)$, $b_2=3.4(1)$, with a $\chi^2$/dof= 21. The fit is shown on the left plot of Fig.~\ref{lnfits}. We note that the fit does seems to roughly pass through all the points although the the $\chi^2$/dof is large. The reason for this is that our data is quite precise and we are sensitive to corrections to scaling assuming they are present. Indeed, if we introduce corrections to scaling and assume
\begin{equation}
 R_{a,p}= \frac{b_a}{L^{1+\eta}} + \frac{c_a}{L^{1 +\eta + \omega }} 
\label{rpfitcorr}
\end{equation}
and fix for example $\omega = 1$ then one can fit the entire data set ($24 \le L \le 44$) for both correlation ratios very well. We obtain $b_1 =7.92(8)$, $c_1=20(2)$, $b_2=3.91(2)$, $c_2=-19.2(6)$, with a $\chi^2$/dof $= 1.1$. The goodness of the fit is shown in the right plot of Fig.~\ref{lnfits}.

\begin{figure*}[htb!]
\vskip0.2in
\includegraphics[width=0.45\textwidth]{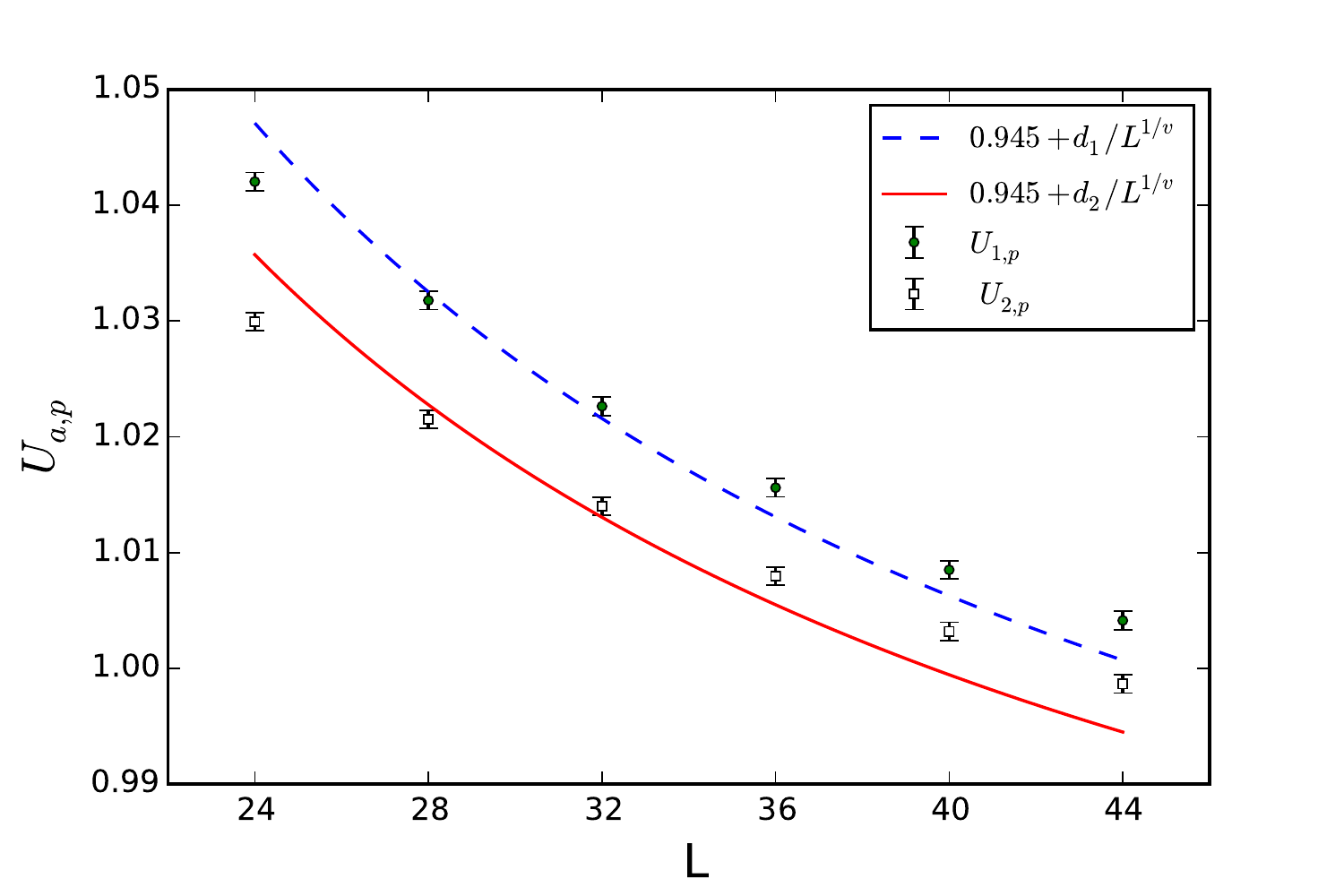}
\includegraphics[width=0.45\textwidth]{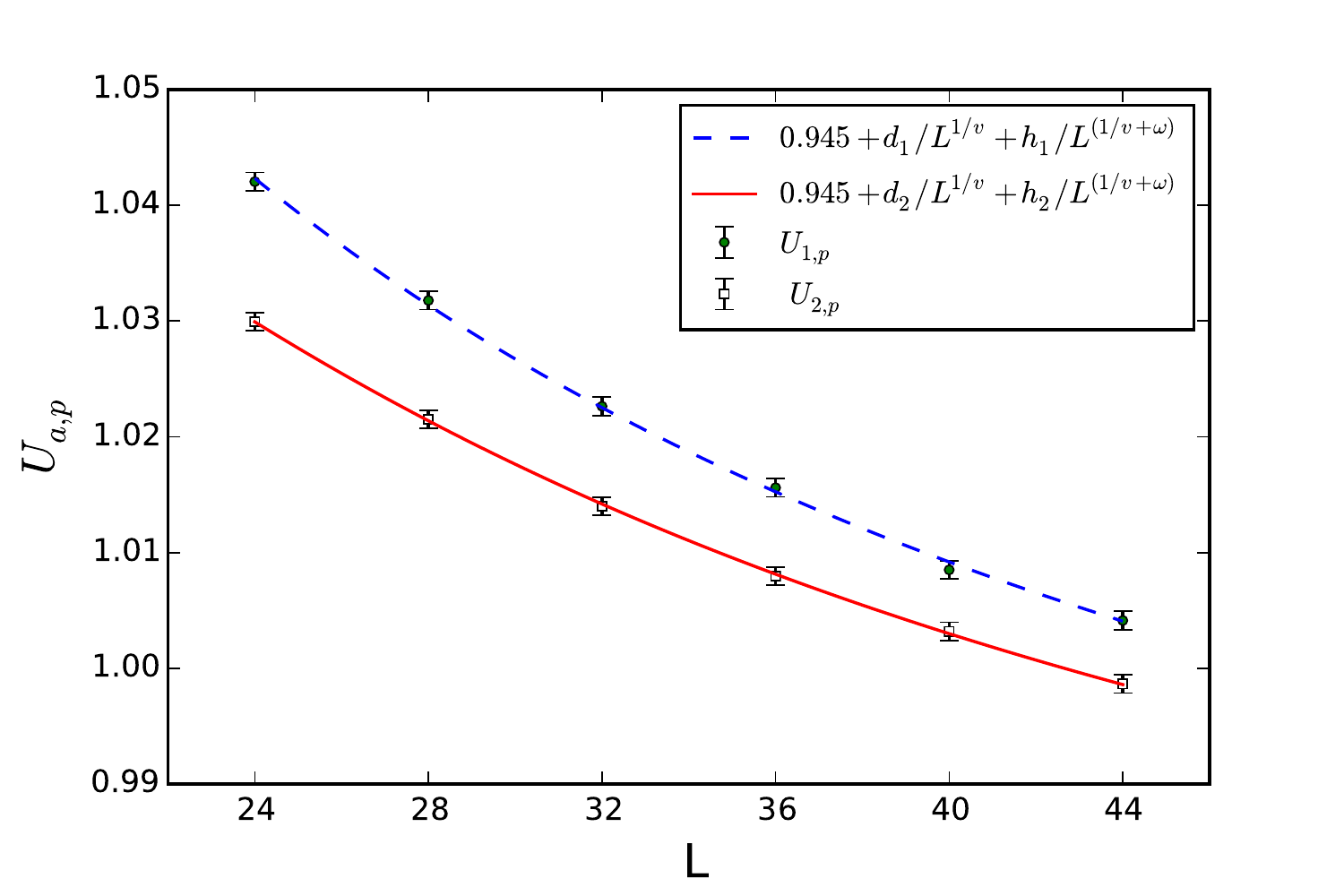}
\caption{\label{lnnufits} Plots of $U_{1,p}$, $U_{2,p}$ as a function of $L$ showing fits to large $N$ exponents with (right) and without (left) corrections to scaling. In the fits we fix $U_c=0.945$, $\nu=1$ and $\omega=1$. The $\chi^2$/dof for the fits are $20$ when the corrections to scaling are omitted and $0.2$ when they are included.}
\end{figure*}

From table \ref{eta_Uc_table_2}, we note that $\eta =1$ gives $U_c=0.945$ assuming the corrections to scaling are small at the critical point. Fixing $ \nu =1$ and $ U_c=0.945$, a combined fit of $U_{1,p}$ and $U_{2,p}$ data to the form (\ref{nup}) gives $d_1= 2.45(5)$, $d_2=2.18(5)$ with a $\chi^2$/dof $= 20$. This is clearly a bad fit as shown in the left plot of Fig.~\ref{lnnufits}.  On the other hand if we introduce corrections to scaling and assume
\begin{equation}
 U_{a,p} = U_c + \frac{d_a}{L^{\frac{1}{\nu}}} + \frac{h_a}{L^{\left(\frac{1}{\nu}+ \omega \right)}},
\label{upfitcorr}
\end{equation}
with $U_c = 0.945$, $\nu=1$ and $\omega=1$ as before, we obtain a good fit as shown in the right plot of Fig \ref{lnnufits}. The fit yields $d_1= 2.91(2)$, $h_2= -13.9(7)$, $d_2=2.74(2)$, $h_2= -16.9(7)$ which a $\chi^2$/dof $= 0.2$. If we do not fix $U_c$ while ignoring the corrections to scaling, again we obtain a good fit with $U_c=0.960(1)$, $d_1= 1.98(4)$, $d_2=1.70(4)$, and $\chi^2$/dof $= 1.18$. However, this value of $U_c$ cannot be consistent with our data in table \ref{eta_Uc_table_2}, again suggesting the presence of large corrections to scaling. 

Thus, we believe that including scaling corrections will enable us to fit the data to large $N$ exponents of $\eta = 1$ and $\nu =1$. However, if we take this view point one has to argue that there are significant corrections to scaling even up to $L=44$. On the other hand since we were able to fit the data without corrections to scaling to a different set of exponents, it is likely that our original analysis may in fact be correct.

\section{Universal Function}

\begin{figure*}[htb!]
\hfill
\includegraphics[width=0.45\textwidth]{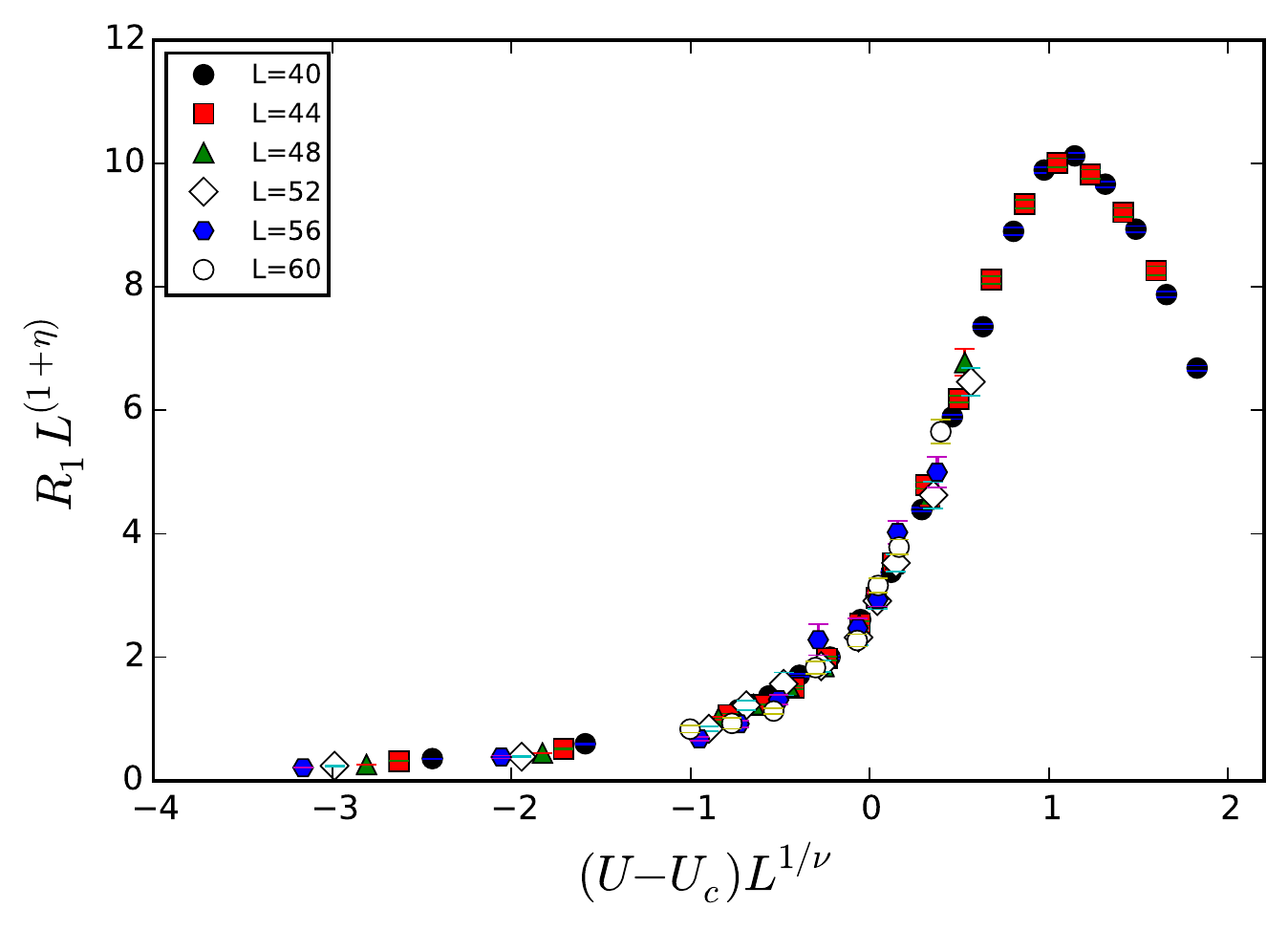}
\hfill
\includegraphics[width=0.45\textwidth]{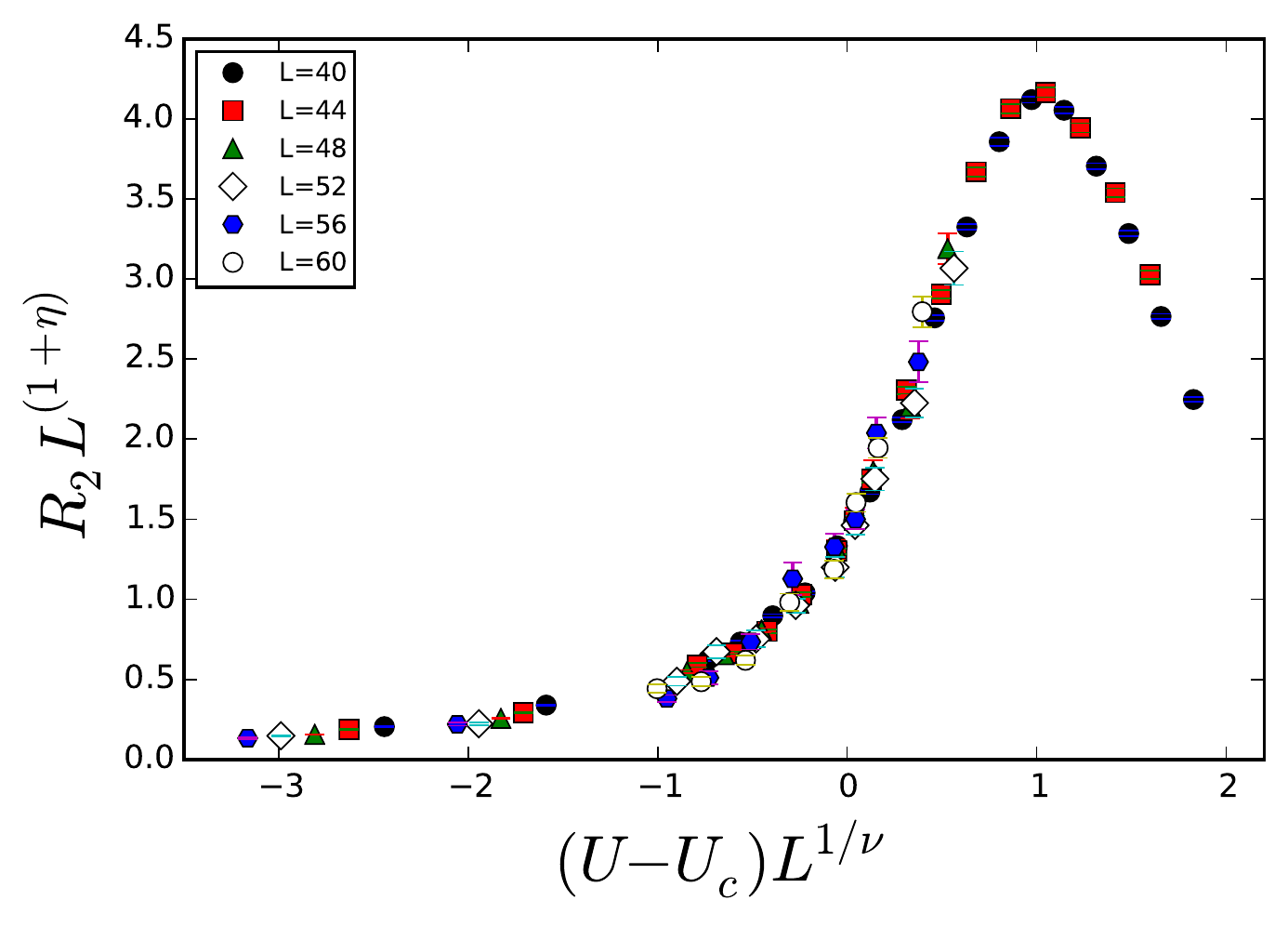}
\hfill
\caption{\label{univ1} Plots of the universal functions $g_1(x)$ (left) and $g_2(x)$ (right) assuming $ U_c = 0.943$,  $\nu=1.30$, and $\eta= 1.05$}
\end{figure*}

\begin{figure*}[htb!]
\hfill
\includegraphics[width=0.45\textwidth]{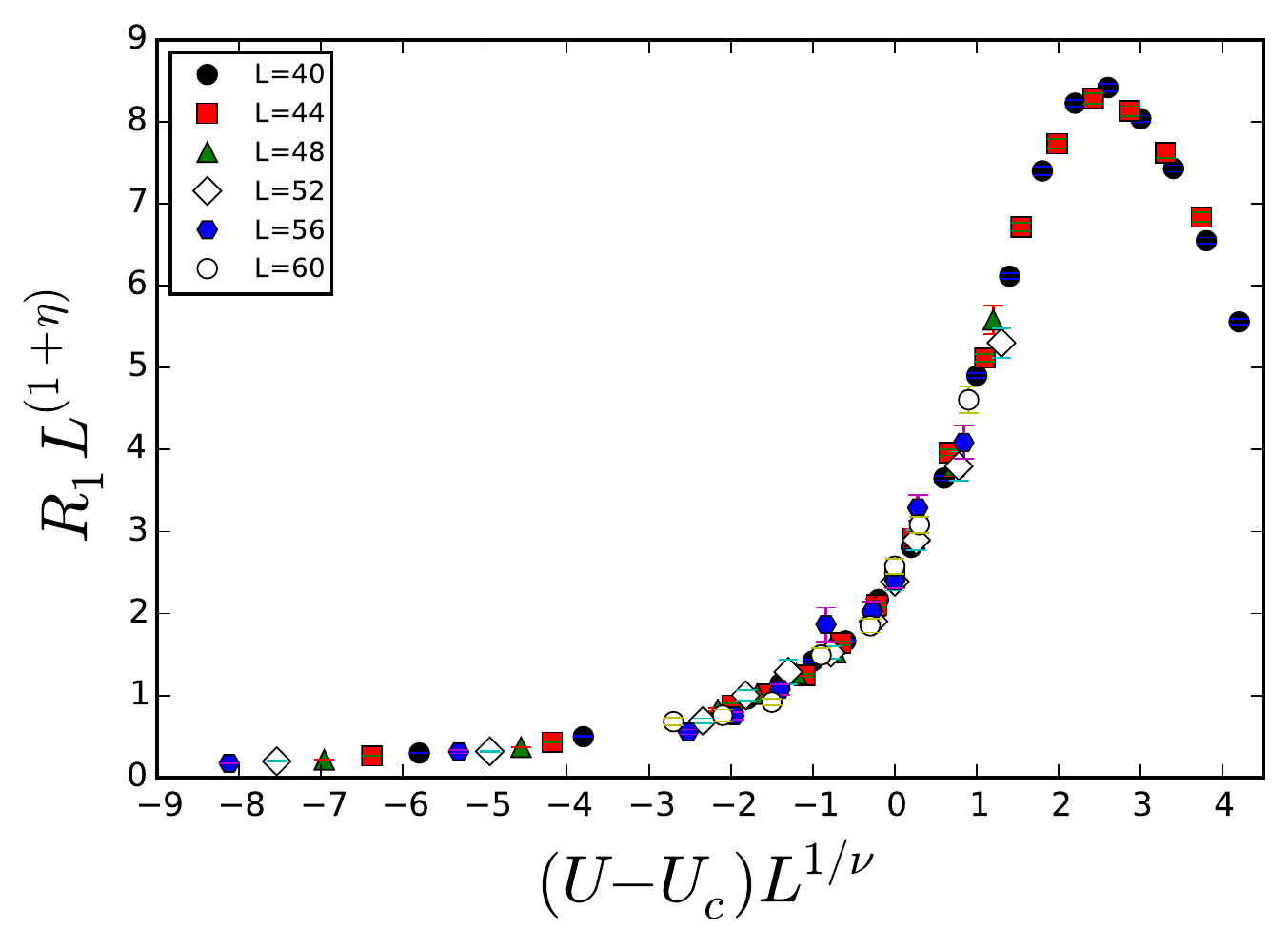}
\hfill
\includegraphics[width=0.45\textwidth]{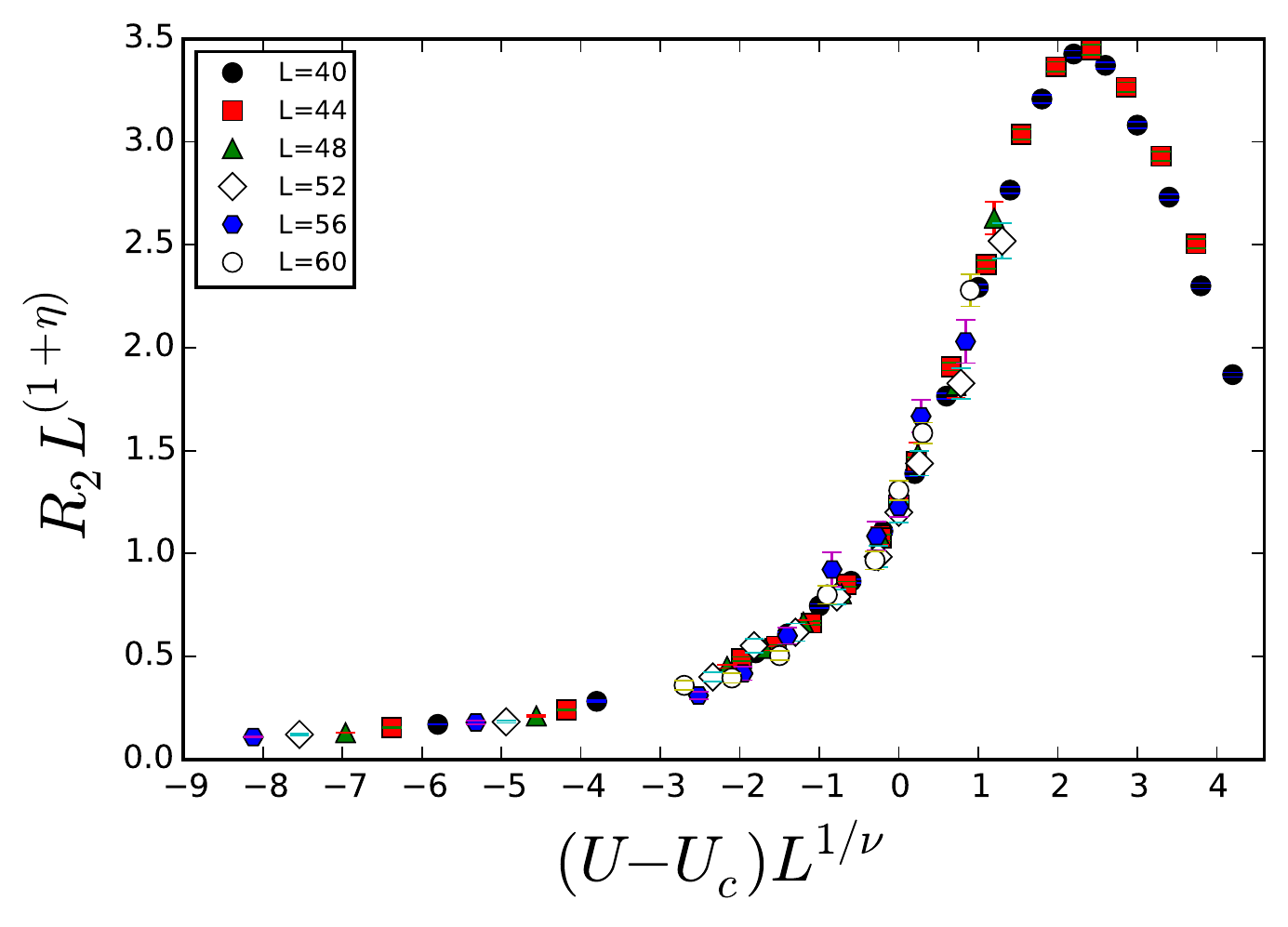}
\hfill
\caption{\label{univ2} Plots of the universal functions $g_1(x)$ (left) and $g_2(x)$ (right) assuming $ U_c=0.945$, $\nu=1$, and $\eta=1$.}
\end{figure*}

In order to contrast the analysis that ignored corrections to scaling and that which included the corrections to scaling but fixed $\eta$ and $\nu$ to their large $N$ values, we plot the universal function $f_a(x)$ introduced in (\ref{scal}). For this purpose we plot $R_a L^{1+\eta} $ as a function of $(U-U_c) L^{\frac{1}{\nu}}$. The plot using $U_c=0.943$, $\eta=1.05$ and $\nu=1.30$ is shown in Fig. \ref{univ1}, while the plot using $U_c=0,945$, $\eta=1$ and $\nu=1$ is shown in  in Fig. \ref{univ2}. To the eye both curves look reasonable, although as explained above significant corrections to scaling are required before the data can be fit to the latter exponents.

Based on the above analysis we can conclude that either we have a new set of exponents with $\eta=1.05(5)$ and $\nu = 1.30(7)$, or there are large corrections to scaling up to lattice sizes of the order of $L=44$ and the exponents are very close to the large $N$ values.

While calculations at larger lattice may be useful to get better estimates of the critical exponents, given the difficulty in performing large scale Monte Carlo calculations it will be useful to explore a new technique of analysis that reduces the systematic errors due to corrections to scaling. 

\end{document}